\documentclass[]{article}
\usepackage{lmodern}
\usepackage{amssymb,amsmath}
\usepackage[utf8]{inputenc}
\usepackage{amsthm}
\usepackage{bbm}
\usepackage{amsfonts}
\usepackage{float}
\usepackage{subfig}
\usepackage{siunitx}
\usepackage{graphicx}
\usepackage{epstopdf}
\usepackage{multirow}
\usepackage{caption}
\usepackage{xcolor}
\theoremstyle{plain}
\newtheorem{theorem}{Theorem}[section]

\newtheorem{proposition}[theorem]{Proposition}

\theoremstyle{definition}

\theoremstyle{remark}

\usepackage[margin=1.2in]{geometry}
\usepackage{color}
\usepackage{fancyvrb}

\DefineVerbatimEnvironment{Highlighting}{Verbatim}{commandchars=\\\{\}}
\makeatletter
\def\maxwidth{\ifdim\Gin@nat@width>\linewidth\linewidth\else\Gin@nat@width\fi}
\def\maxheight{\ifdim\Gin@nat@height>\textheight\textheight\else\Gin@nat@height\fi}
\makeatother
\setkeys{Gin}{width=\maxwidth,height=\maxheight,keepaspectratio}
\makeatletter
\def\fps@figure{htbp}
\makeatother
\title{Finite Mixture Approximation of CARMA(p,q) Models}
\author{Lorenzo Mercuri, Andrea Perchiazzo, Edit Rroji}

\begin{document}
\maketitle
\begin{abstract}
In this paper we show how to approximate the transition density of a
CARMA(p, q) model driven by means of  a time changed Brownian Motion based on the
Gauss-Laguerre quadrature.  We then provide an analytical formula for option prices when the log price
follows a CARMA(p, q) model. We also propose an estimation procedure
based on the approximated likelihood density. 

\end{abstract}

\section{Introduction} \label{introduction}

The aim of this paper is to provide a simple approximation procedure for the transition density of a Continuous Autoregressive Moving Average Model driven by a Time Changed Brownian Motion. The Continuous Autoregressive Moving Average (CARMA hereafter) model with guassian transition density was first introduced in \cite{doob1944} as a continuous counterpart of the well known ARMA process defined in discrete time. Recently this model has gained a significant attention in literature due to the relaxation of the gaussianity assumption. 

A L\'evy CARMA model has been proposed in \cite{Brockwell2001} and the associated marginal distribution is allowed to be skewed and fat-tailed. These features increase the appealing of these processes especially in modeling financial time series [see for examples \cite{Brockwell2011, Iacus2015} and references therein]. Indeed, in  CARMA(p,q) models it is possible to work directly with market data without being forced of considering an equally spaced time grid necessary in discrete-time models like for example in ARMA(p,q) models. 
 
The CARMA(p,q) process can be seen as a generalization of the Ornstein-Uhlenbeck process (OU). The  OU process is not sufficiently flexible for financial applications since its autocorrelation function shows a monotonic decreasing (negative exponential) behaviour. In this context, the CARMA(p,q) model  seems to be useful as it is able to capture a more complex shape for the dependence structure as discussed in \cite{Brockwell2004}. The nice statistical and mathematical properties make this class of continuous time models very suitable for modeling commodities \cite{NUALART2000,BENTH2014392}, interest rates \cite{andresen2014carma}, mortality intensity \cite{Hitaj2019}, spot electricity prices \cite{garcia2011estimation} and temperature \cite{benth2007putting}. 

In order to apply the CARMA model on real data, for the evaluation of derivatives on commodities and/or for the evaluation of insurance contracts, it is necessary to know the transition density of the process. In the case of a CARMA(p,q) model where the driving noise is a Brownian motion, the  transition density is Gaussian. Therefore, an estimation procedure [see \cite{Tomasson2015} for details] can be obtained directly combining the Gaussian likelihood function with the Kalman Filter while for the pricing of financial/insurance contracts we have to compute just the expected value of a transformation of a Gaussian  random variable. We refer for instance to the pricing formula for options on futures derived in \cite{PASCHKE20102742} where the log-spot price is a gaussian CARMA(p,q) process. Similar results are obtained for interest rate derivatives [see \cite{andresen2014carma} for details].

The main contribution of this paper is to propose a finite mixture of normals that approximates the transition density of a Time Changed Brownian Motion CARMA(p,q) process (TCBm-CARMA hereafter). This approximation increases the appealing of the CARMA model in practical applications since, as a finite mixture of normals, it has a level of computational complexity similar to the gaussian CARMA for estimation on real data and for evaluation of financial and insurance contracts. The choice of a Time Changed Brownian Motion (TCBm) as a driving noise increases also the ability of the CARMA to capture the statistical features of data. In the case of the TCBm-CARMA,  our results generalize in a straightforward manner the estimation procedure in \cite{Iacus2015} based on the Quasi-Gaussian Likelihood (QGMLE) contrast function [see \cite{yoshida2011polynomial,Masuda2013} and reference therein for a complete discussion of the QGMLE procedure]. Indeed we do not need a two step procedure but we are able to estimate autoregressive, moving average and L\'evy measure parameters at the same time.  Pricing formulas of financial contracts are again simple linear convex combinations of gaussian pricing formulas. For instance for options written on futures we have a convex linear combination of pricing formulas in \cite{PASCHKE20102742}. 

The construction of our approximated transition density for a TCBm-CARMA(p,q) model is based on two main components: the dyadic Riemann sum approximation of a stochastic integral [see \cite{attal2003quantum} for a complete discussion] and the Gauss-Laguerre quadrature [see \cite{abramowitz70a} for more details]. The main idea behind this approach is to approximate the distribution associated to the subordinator process at unitary time with a discrete random variable where the realizations are the zeros of the Laguerre polynomial with a fixed order and the corresponding probability is obtained using the Gauss-Laguerre quadrature. \newline Based on our knowledge the first authors that applied this approach in two different situation are \cite{Madan2013} for a option pricing purpose and \cite{LOREGIAN2012217} for the estimation of the Variance Gamma distribution using the EM-algorithm proposed by \cite{Dempster77maximumlikelihood}. Several authors, recently have used the Laguerre polynomials to derive approximated closed formulas for the pricing of financial contracts [see \cite{Belle2019} and reference therein] and insurance contracts [see \cite{ZHANG2019329} and reference therein] for some specific exponential L\'evy processes. A comparison of some numerical techniques including the Gauss-Laguerre quadrature for pricing derivatives  under an exponential Variance Gamma process has been presented in \cite{Aguilar2020}.

The paper is organized as follows. Section \ref{finite-approximation-of-the-density-of-a-normal-variance-mean-mixture} reviews the Gauss-Laguerre approximation for a Normal Variance Mean Mixture random variable. In Section \ref{LCARMA} we extend the Gauss-Laguerre approximation to the case of the transition density of a TCBm-CARMA(p,q) model and we propose an estimation method that maximizes the approximated likelihood function. In Section \ref{option-pricing-in-a-luxe9vy-carmapq-model.} we discuss how to apply our approximated density in the evaluation of a transformation of the exponential TCBm-CARMA(p,q) model. In particular we derive specific formulas for the futures term structure and for option prices on futures. Section \ref{conclusion} concludes the paper.

\section{Finite Approximation of the Density of a Normal Variance Mean Mixture} \label{finite-approximation-of-the-density-of-a-normal-variance-mean-mixture}

First we recall the formal definition of a Normal Variance Mean Mixture
discussed in \cite{Barndorff97}. A random
variable \(Y\) is a Normal Variance Mean Mixture if we have:
\begin{equation}
Y = \mu + \theta \Lambda + \sigma\sqrt{\Lambda}Z
\label{1}
\end{equation} \(Z\sim N\left(0,1\right)\). \(\Lambda\) is a continuous
positive random variable with an exponentially slowly density function
\(f\left(u\right)\) defined as: \begin{equation}
f\left(u\right)=e^{-\varphi_+u}u^{\lambda-1}L_{\theta}\left(u\right)\mathbbm{1}_{\left\{u\geq0\right\}},
\label{2}
\end{equation} \(\varphi_+\geq0\),
\(L_{\theta}\left(u\right):\left[0,+\infty\right)\rightarrow\left[0,+\infty\right)\)
function with slowly variation, i.e.: \[
\lim_{u\rightarrow+\infty}\frac{L\left(\alpha u\right)}{L\left(u\right)}=1.
\] In order to construct a discrete version of the random variable
\(\Lambda\), we use the Gauss-Laguerre quadrature. Let \(f\left(x\right)\) be
a function with support \(\left[0,+\infty\right)\) such that \[
\int_0^{+\infty} f\left(x\right)e^{-x}\mbox{d}x<+\infty,
\] we have the follwing approximation: \begin{equation}
\int_0^{+\infty} f\left(x\right)e^{-x}\mbox{d}x\approx\sum_{i=1}^{m}w\left(k_i\right)f\left(k_i\right).
\label{approx:Laguerre}
\end{equation} \(k_i\) is the \(i\)-th root of the Laguerre polynomial
\(L_m\left(k_i\right)\) and the weights
\(w\left(k_i\right), \ i=1,\ldots,m\) are: \[
w\left(k_i\right) = \frac{k_i}{\left(m+1\right)^2L^2_{m+1}\left(k_i\right)}.
\] We start from the moment generating function of the random variable
\(\Lambda\): \begin{equation}
\mathbb{E}\left(e^{c\Lambda}\right)=\int_0^{+\infty}e^{cu}e^{-\varphi_+u}u^{\lambda-1}L_{\theta}\left(u\right)\mbox{d}u.
\label{mgf}
\end{equation} Posing \(k=\varphi_+u\) in \eqref{mgf}, we get: \[
\mathbb{E}\left(e^{c\Lambda}\right)=\int_0^{+\infty}e^{-k}e^{c\frac{k}{\varphi}}\left(\frac{k}{\varphi}\right)^{\lambda-1}L_{\theta}\left(\frac{k}{\varphi}\right)\frac{\mbox{d}k}{\varphi}=\int_0^{+\infty}e^{-k}\frac{e^{c\frac{k}{\varphi}}}{k}\left(\frac{k}{\varphi}\right)^{\lambda}L_{\theta}\left(\frac{k}{\varphi}\right)\mbox{d}k.
\] Applying the formula in \eqref{approx:Laguerre}, we have: \[
\mathbb{E}\left(e^{c\Lambda}\right)\approx \sum_{i=1}^m e^{c\left(\frac{k_i}{\varphi_+}\right)}\frac{w\left(k_i\right)}{k_i}\left(\frac{k_i}{\varphi_+}\right)^{\lambda}L_{\theta}\left(\frac{k_i}{\varphi_+}\right).
\] It is to worth noting that \[
\sum_{i=1}^m \frac{w\left(k_i\right)}{k_i}\left(\frac{k_i}{\varphi_+}\right)^{\lambda}L_{\theta}\left(\frac{k_i}{\varphi_+}\right)\approx\int_0^{+\infty}\frac{e^{-k}}{k}\left(\frac{k}{\varphi_+}\right)^{\lambda}L_{\theta}\left(\frac{k}{\varphi_+}\right)\mbox{d}k.
\] Using the substitution \(u=\frac{k}{\varphi_+}\) we get: \[
\sum_{i=1}^m \frac{w\left(k_i\right)}{k_i}\left(\frac{k_i}{\varphi_+}\right)^{\lambda}L_{\theta}\left(\frac{k_i}{\varphi_+}\right)\approx\int_0^{+\infty}e^{-\varphi_+u}u^{\lambda-1}L_\theta\left(u\right)\mbox{d}u=1,
\] therefore we have: \begin{equation}
\mathbb{E}\left(e^{c\Lambda}\right)\approx \sum_{i=1}^m e^{c\left(\frac{k_i}{\varphi_+}\right)}\frac{\frac{w\left(k_i\right)}{k_i}\left(\frac{k_i}{\varphi_+}\right)^{\lambda}L_{\theta}\left(\frac{k_i}{\varphi_+}\right)}{\sum_{i=1}^m \frac{w\left(k_i\right)}{k_i}\left(\frac{k_i}{\varphi_+}\right)^{\lambda}L_{\theta}\left(\frac{k_i}{\varphi_+}\right)}.
\label{eq:MgfApprox}
\end{equation} The right hand side of the equation \eqref{eq:MgfApprox}
can be seen as the moment generating function of a positive random
variable \(\Lambda_m\) with a finite support and defined as: \begin{equation}
\Lambda_m=\left\{
\begin{array}{lcl}
u_1=\frac{k_1}{\varphi_+}& & \mathbb{P}\left(u_1\right)=\frac{\frac{w\left(k_1\right)}{k_1}\left(\frac{k_1}{\varphi_+}\right)^{\lambda}L_{\theta}\left(\frac{k_1}{\varphi_+}\right)}{\sum_{i=1}^m \frac{w\left(k_i\right)}{k_i}\left(\frac{k_i}{\varphi_+}\right)^{\lambda}L_{\theta}\left(\frac{k_i}{\varphi_+}\right)}\\
\vdots & & \vdots\\
u_i=\frac{k_i}{\varphi_+}& & \mathbb{P}\left(u_i\right)=\frac{\frac{w\left(k_i\right)}{k_i}\left(\frac{k_i}{\varphi_+}\right)^{\lambda}L_{\theta}\left(\frac{k_i}{\varphi_+}\right)}{\sum_{i=1}^m \frac{w\left(k_i\right)}{k_i}\left(\frac{k_i}{\varphi_+}\right)^{\lambda}L_{\theta}\left(\frac{k_i}{\varphi_+}\right)}\\
\vdots & & \vdots\\
u_m=\frac{k_m}{\varphi_+}& & \mathbb{P}\left(u_m\right)=\frac{\frac{w\left(k_n\right)}{k_m}\left(\frac{k_m}{\varphi_+}\right)^{\lambda}L_{\theta}\left(\frac{k_m}{\varphi_+}\right)}{\sum_{i=1}^m \frac{w\left(k_i\right)}{k_i}\left(\frac{k_i}{\varphi_+}\right)^{\lambda}L_{\theta}\left(\frac{k_i}{\varphi_+}\right)}\\
\end{array}
\right. .
\label{eq:approxMixRVNVMM}
\end{equation} The next step is to consider a sequence of random
variables \(Y_m\) defined as: \begin{equation}
Y_m=\mu+\theta \Lambda_m+\sqrt{\Lambda_m}Z,
\end{equation} with \(Z\sim N\left(0,1\right)\) independent of \(\Lambda_m\).
For any \(m\) the density of \(Y_m\) is a finite mixture of normal with
the following form: \begin{equation}
f_{Y_m}\left(y\right)=\sum_{i=1}^m\phi(y,\mu_0+\mu u_i,\sigma^2 u_i)\mathbb{P}\left(u_i\right)
\label{approxDens}
\end{equation} where \(\phi(x,a,b)\) is a normal density at point \(x\)
with mean \(a\) and variance \(b\). Using the definition of \(\Lambda_m\)
\begin{equation}
f_{Y_m}\left(y\right)=\sum_{i=1}^m\phi\left(y,\mu+\theta \frac{k_i}{\varphi_+};\sigma^2 \frac{k_i}{\varphi_+}\right)\frac{\frac{w\left(k_i\right)}{k_i}\left(\frac{k_i}{\varphi_+}\right)^{\lambda}L_{\theta}\left(\frac{k_i}{\varphi_+}\right)}{\sum_{i=1}^m \frac{w\left(k_i\right)}{k_i}\left(\frac{k_i}{\varphi_+}\right)^{\lambda}L_{\theta}\left(\frac{k_i}{\varphi_+}\right)}
\label{eq:approximated}
\end{equation} Applying the Gauss-Laguerre quadrature we get: \[
f_{Y_m}\left(y\right)\stackrel{m\rightarrow+\infty}{\longrightarrow}\int_0^{+\infty}\phi\left(y,\mu_0+\mu \frac{k}{\varphi_+}; \sigma^2 \frac{k}{\varphi_+}\right)\frac{e^{-k}}{k}\left(\frac{k}{\varphi_+}\right)^{\lambda}L_{\theta}\left(\frac{k}{\varphi_+}\right)\mbox{d}k.
\] Substituting \(u=\frac{k}{\varphi_+}\), we have: \[
f_{Y_m}\left(y\right)\stackrel{m\rightarrow+\infty}{\longrightarrow}\int_0^{+\infty}\phi\left(y,\mu_0+\mu u;\sigma^2 u\right)e^{-\varphi_+u}u^{\lambda-1}L_{\theta}\left(u\right)\mbox{d}u.
\] The right-hand side is the density of the random variable in
\eqref{1}. Observe that approximation discussed here can be applied in
three wide applied distributions: Variance Gamma, Normal Inverse
Gaussian, Generalized Hyperbolic. In all cases, the density of the
mixing random variable belongs to the class defined in \eqref{2}. Indeed
we obtain the density of a Gamma random variable with shape \(\alpha\)
and rate \(\beta\) parameters posing the following condition: \[
\varphi_+=\beta,\ \lambda = \alpha, \ L_{\left(\alpha,\beta\right)}\left(u\right)=\frac{\beta^{\alpha}}{\Gamma\left(\alpha\right)},
\] therefore the density in \eqref{approxDens} approximate the density
of a Variance Gamma random variable.\newline The density of an Inverse
Gaussian IG\(\left(a,b\right)\) can be obtained from \eqref{1} by
posing: \[
\varphi_+=\frac{b^2}{2},\ \lambda=-\frac12, \ L_{a,b}\left(u\right)=\left[\frac{a}{\sqrt{2\pi}}\right]e^{ab-\frac{a^2}{2x}}.
\] In this case we obtain an approximation of the Normal Inverse
Gaussian density using \eqref{approxDens}.\newline The Generalized
Inverse Gaussian density with \(a>0\), \(b>0\) and \(p\in\mathbb{R}\) is
a special case of \eqref{1} when: \[
\varphi_+=\frac{\alpha}{2},\ \lambda=p, \ L_{a,b,p}\left(u\right)=\frac{\left(\frac{a}{b}\right)^{\frac{p}{2}}}{2K_p\left(\sqrt{ab}\right)}e^{-\frac{b}{2u}}
\] where \(K_p\left(x\right)\) is a modified Bessel function of the
second kind. Using \eqref{approxDens} we approximate the density of a
Generalized Hyperbolic distribution.

\begin{figure}[!htbp]
\centering
\includegraphics[width=0.5\textwidth]{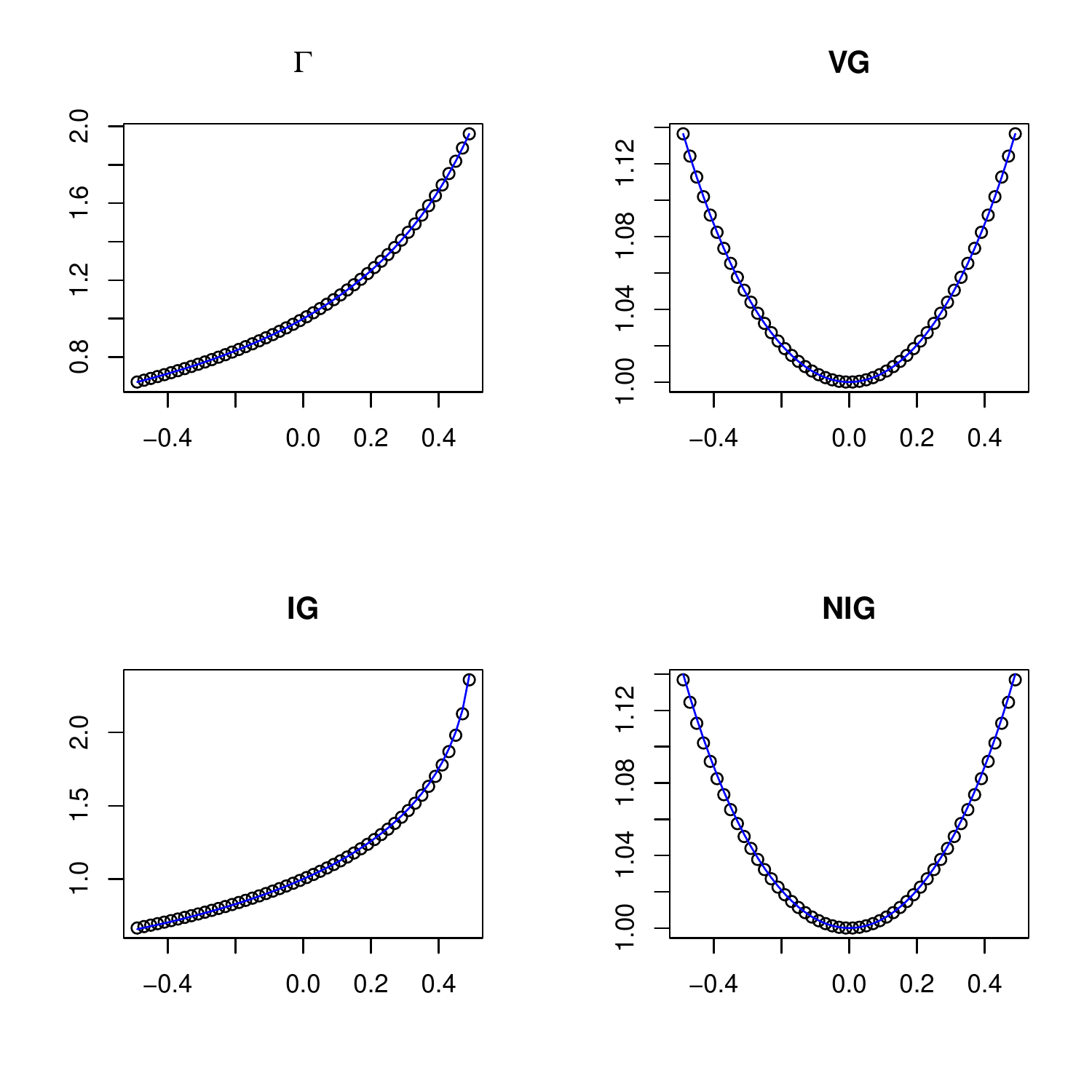}
\caption{\label{fig:Fig1} Comparison between theoretical and
approximated moment generating function for a
\(\Gamma\left(1,1\right)\), the corresponding symmetric Variance Gamma
centered in zero, a IG\(\left(1,1\right)\) and its associated symmetric
Normal Inverse Gaussian centered in zero.}
\end{figure}

Figure \ref{fig:Fig1} shows the behavior of the analytic and
approximated moment generating functions for the Gamma, Variance Gamma,
Inverse gaussian, Normal Inverse Gaussian model. To generate the
approximated moment generating function we use \(m=40\).

In Appendix \ref{EMderivation} derive the Expectation Maximization algorithm for the approximated
density in \eqref{eq:approximated}.

\section{L\'evy CARMA(p,q) model.} \label{LCARMA}

In this section, we review the main features of L\'evy CARMA(p,q) models.
The CARMA model, firstly introduced by \cite{doob1944} as a generalization
in continuous time setup of the Gaussian ARMA model, has recently
gained a rapid development in different areas due to the substitution
of the Brownian Motion with a general L\'evy process as driving noise [see \cite{Brockwell2001} for a discussion of a CARMA model driven by a L\'evy process with finite second order moments].

The formal definition of a L\'evy CARMA(p,q) model \(Y_t\) with
\(p > q \geq 0\) is based on the continuous version of the state-space
representation of an autoregressive moving average-ARMA(p,q) model:

\begin{equation}
Y_t =\mathbf{b}^{\top}X_t
\label{eqCar}
\end{equation}

where \(X_t\) satisfies: \begin{equation}
\mbox{d}X_t = \mathbf{A}X_{t-}\mbox{d}t+\mathbf{e}\mbox{d}Z_t.
\label{eq:CarSDE}
\end{equation} $\left\{Z_{t}\right\}_{t\geq0}$ is a L\'evy process. The matrix \(\mathbf{A}\) with dimension \(p\times p\) is
defined as: \begin{equation*}
\mathbf{A}=\left[
\begin{array}{ccccc}
\\
0 & 1 & 0 & \ldots & 0\\
0 & 0 & 1 & \ldots & 0\\
\vdots & \vdots & \vdots & \ddots & \vdots\\
0 & 0 & 0 & \ldots & 1\\
-a_p & -a_{p-1} & -a_{p-2} & \ldots & -a_1\\
\end{array}
\right]_{p\times p}.
\end{equation*} The vectors \(\mathbf{e}\) and \(\mathbf{b}\) with
dimension \(p\times 1\) are defined as follows: \[
\mathbf{e}=\left[0,0,\ldots,1\right]^\top
\] \[
\mathbf{e}=\left[b_0,0,\ldots,b_{p-1}\right]^\top
\] where \(b_{q+1}=\ldots=b_{p-1}=0\). Given the initial point \(X_s\),
the solution of thee Eq. \eqref{eq:CarSDE} is: \[
X_t= e^{\mathbf{A}\left(t-s\right)}X_s+\int_{0}^{+\infty}e^{\mathbf{A}\left(t-s\right)}\mbox{d}Z_u, \ \forall t>s,
\] where
\(e^\mathbf{A}=\underset{h=0}{\stackrel{+\infty}{\sum}}\frac{1}{h!}\mathbf{A}^h\).\newline
We report in the following the scale property of a CARMA(p,q) process.
This property introduces a constraint between the L\'evy measure
parameters and the moving average vector \(\mathbb{b}\). Indeed it is
possible to introduce a new L\'evy process \(L_t\) defined as: \[
L_t=\frac{1}{a}Z_t, \ a>0.
\] We also define the state process \(X^{\prime}_t\) as: \[
X^{\prime}_t =\frac{1}{a}X_t
\] and a new moving average vector \(\tilde{\mathbf{b}}=a\mathbf{b}\),
the CARMA(p,q) process in \eqref{eqCar} can be written equivalently as: \[
Y_t= \tilde{\mathbf{b}}^\top X^{\prime}_t
\] where \(X^{\prime}_t\) satisfies the following Stochastic
Differential Equations: \[
\mbox{d}X_t^{\prime}=A X_{t-}^{\prime}\mbox{d}t+\mathbf{e}\mbox{d}L_t.
\] As reported in \cite{Brockwell2011}, under the
assumption that all eigenvalues \(\lambda_1,\ldots,\lambda_p\) of matrix
\(\mathbf{A}\) are distinct and their real part is negative, we can
write the CARMA(p,q) model as a summation of a finite number of
continuous autoregressive models of order 1, i.e.~CAR(1) models.
Therefore: \begin{equation}
Y_t = \mathbf{b}^\top e^{\mathbf{A}\left(t-s\right)} X_s +\int_0^{+\infty}\underset{i=1}{\stackrel{p}\sum}\left[\alpha\left(\lambda_i\right)e^{\lambda_i\left(t-u\right)}\right]\mathbb{I}_{s\leq u\leq t}\mbox{d}Z_u 
\label{expr:Sol}
\end{equation} with
\(\alpha\left(z\right)=\frac{b\left(z\right)}{a^{\prime}\left(z\right)}\)
where \(a\left(z\right)\) and \(b\left(z\right)\) are polynomial
functions defined as: \[
a\left(z\right)=z^p+a_1z^{p-1}+\ldots+a_p,
\] \[
b\left(z\right)=b_0+b_1z+\ldots+b_{p-1}z^{p-1}.
\] Under the additional requirement of the existence of a cumulant
generating function for \(Z_1\), the conditional moment generating
function of a CARMA(p,q) model \(Y_t\) given the information at time
\(s<t\) is obtained: \begin{equation}
\mathbb{E}_s\left[e^{cY_t}\right]=e^{c\mathbf{b}^\top e^{\mathbf{A}\left(t-s\right)X_s}}\exp\left[\int_{s}^{t}\kappa\left(c\underset{i=1}{\stackrel{p}{\sum}}\left[\alpha\left(\lambda_i\right)e^{\lambda_i\left(t-u\right)}\right]\right)\mbox{d}u\right]
\label{mgfCARMA}
\end{equation} with
\(\kappa\left(c\right)=\ln \mathbb{E}\left[e^{cZ_1}\right]<+\infty.\)
\newline Once the state variable \(X_s\) is filtered from observable data, from a
theoretical point of view, the result in \eqref{mgfCARMA} can be used to
compute the transition density from time \(s\) to time \(t\) by means of
the Fourier Transform because the characteristic function is obtained
from the moment generating function evaluated at \(iu\).\newline In
order to get an estimate of the state variable from the observed data
\(Y_{t_0},Y_{t_1},\ldots, Y_{t_i},\ldots\), it is possible to use the
approach discussed in \cite{Brockwell2011} and recently
implemented in \cite{Iacus2015}. As first step, the vector
\(\hat{X}_{q,t}\) containing the first \(q-1\) components of the state
process \(X_t\) can be written in terms of \(Y_{t-1}\) as follows: 
\begin{equation}
\mbox{d}\hat{X}_{q,t} = \mathbf{B}\hat{X}_{q,t-}\mbox{d}t+\mathbf{e}Y_{t-1}\mbox{d}t
\label{eq:StochFilterBrock}
\end{equation} 
where 
\begin{equation*}
\mathbf{B}=\left[
\begin{array}{ccccc}
\\
0 & 1 & 0 & \ldots & 0\\
0 & 0 & 1 & \ldots & 0\\
\vdots & \vdots & \vdots & \ddots & \vdots\\
0 & 0 & 0 & \ldots & 1\\
-b_0 & -b_1 & -b_{2} & \ldots & -b_{q-1}\\
\end{array}
\right]_{p\times p}
\end{equation*} and \[
\mathbf{e}_q=\left[0,\ldots,0,1\right]^{\top}.
\] The remaining \(p-q\) components of \(X_t\) are obtained from the higher order derivatives of the first component \(X_{0,t}\) in the state vector, i.e.:
\(X_t\) with respect to time: \[
X_{j,t}=\frac{\partial^{j-1}X_{0,t}}{\left(\partial  t\right)^{j-1}}, j = q,\ldots,p-1.
\] Combining the approach in \cite{Brockwell2011} with the
result in \eqref{mgfCARMA}, it is possible to introduce an estimation
procedure of the L\'evy CARMA(p,q) model based on the Maximum
Likelihood method. This procedure requires the numerical
evaluation of two integrals, the first in the definition of the moment
generating function \eqref{mgfCARMA} and the second in the
inversion formula of the characteristic function. In this section, we show that in the case of a Time
Changed Brownian motion, we can can approximate the density using the Laguerre polynomials overcomung the numerical integration problems that arise in the standard approach. We
start considering the case of the Ornstein Uhlenbeck that does not
require the estimation of the state process then we move to the general
CARMA(p,q) model.

\subsection{Estimation of an Ornstein Uhlenbeck driven by a Time Changed Brownian Motion.}
\label{Est:OUdrivenTCBM}

Let \(\left(\Omega, \mathcal{F}, \mathbb{F}, \mathcal{P}\right)\) be a filtered
probability space where \(\mathbb{F}=\left(\mathcal{F}_t\right)_{t\geq0}\) is a
filtration, the process \(Y_t\) is a Time Changed Brownian Ornstein-Uhlenbeck (TCBm-OU hereafter) \(Y_t\) satisfies the following stochastic
differential equation: \begin{equation}
\mbox{d}Y_t=-aY_{t-}\mbox{d}t+\mbox{d}W_{\Lambda_t}, \ Y_{t_0}=y_0.
\label{eq:OU}
\end{equation} where \(W_{\Lambda_t}\) is a Brownian Motion stopped by the
subordinator process \(\Lambda_t\). The solution of the SDE in \eqref{eq:OU}
is: \[
Y_t = y_0e^{-a\left(t-t_0\right)}+\int_{t_0}^{t}e^{-a\left(t-u\right)}\mbox{d}W_{\Lambda_{u}}.
\] It is worth noting that the distribution at time 1 of the process
\(W_{\Lambda_t}\) is a Normal Variance Mean Mixture centered in zero. Defining
the \(\sigma\)-field
\(\mathcal{G}_{t_0,t}=\sigma\left(\mathcal{F}_{t_0} \cup \sigma\left(\left\{\Lambda_u\right\}_{u\leq t}\right)\right)\)
with \(t_0<t\), we have: \[
W_{\Lambda_t}-W_{\Lambda_{t_0}}\left|\mathcal{G}_{t_0,t}\right.\sim N\left(0,\Lambda_t-\Lambda_{t_0}\right).
\] The \(\sigma\)-field \(\mathcal{G}_{t_0,t}\) is crucial for the
construction of the approximated transition density of the TCBm-OU
process. 

\begin{proposition}
Given the information associated to the \(\sigma\)-field
\(\mathcal{G}_{t_0,t}\), the conditional distribution for $Y_t$ becomes\footnote{Using the result in \eqref{eq:CondDist} and the interated expected value, we obtain the moment generating function of a TCBm-OU process. : 
\begin{eqnarray*}
\mathbb{E}_{\mathcal{F}_{t_0}}\left[\mathbb{E}\left[e^{cY_t}\left|\mathcal{G}_{t_0,t}\right.\right]\right]&=&e^{cy_0e^{-a\left(t-t_0\right)}}\mathbb{E}_{\mathcal{F}_{t_0}}\left[e^{\frac{c^2}{2}\int_{t_0}^te^{-2a\left(t-u\right)}\mbox{d}\Lambda_u}\right]\nonumber\\
&=& e^{cy_0e^{-a\left(t-t_0\right)}+\int_{t_0}^t \kappa_{\Lambda}\left(\frac{c^2}{2}e^{-2a\left(t-u\right)}\right)\mbox{d}u}.
\end{eqnarray*} 
where \(\kappa_{\Lambda}\left(u\right)=\ln\left[\mathbb{E}\left(e^{u{\Lambda}_1}\right)\right]\). The quantity $e^{cy_0e^{-a\left(t-t_0\right)}+\int_{t_0}^t \kappa_{\Lambda}\left(\frac{c^2}{2}e^{-2a\left(t-u\right)}\right)\mbox{d}u}$ is the moment generating function of an TCBm-OU process and it can be alternatively obtained applying the result in \cite{Eberlein1999}.}:

 \begin{equation}
Y_t\left|\mathcal{G}_{t_0,t}\right.\sim N\left(y_0e^{-a\left(t-t_0\right)}, \int_{t_0}^{t}e^{-2a\left(t-u\right)}\mbox{d}\Lambda_u\right).
\label{eq:CondDist}
\end{equation} 
\end{proposition}

\begin{figure}[!htbp]
\centering
\includegraphics[width=0.5\textwidth]{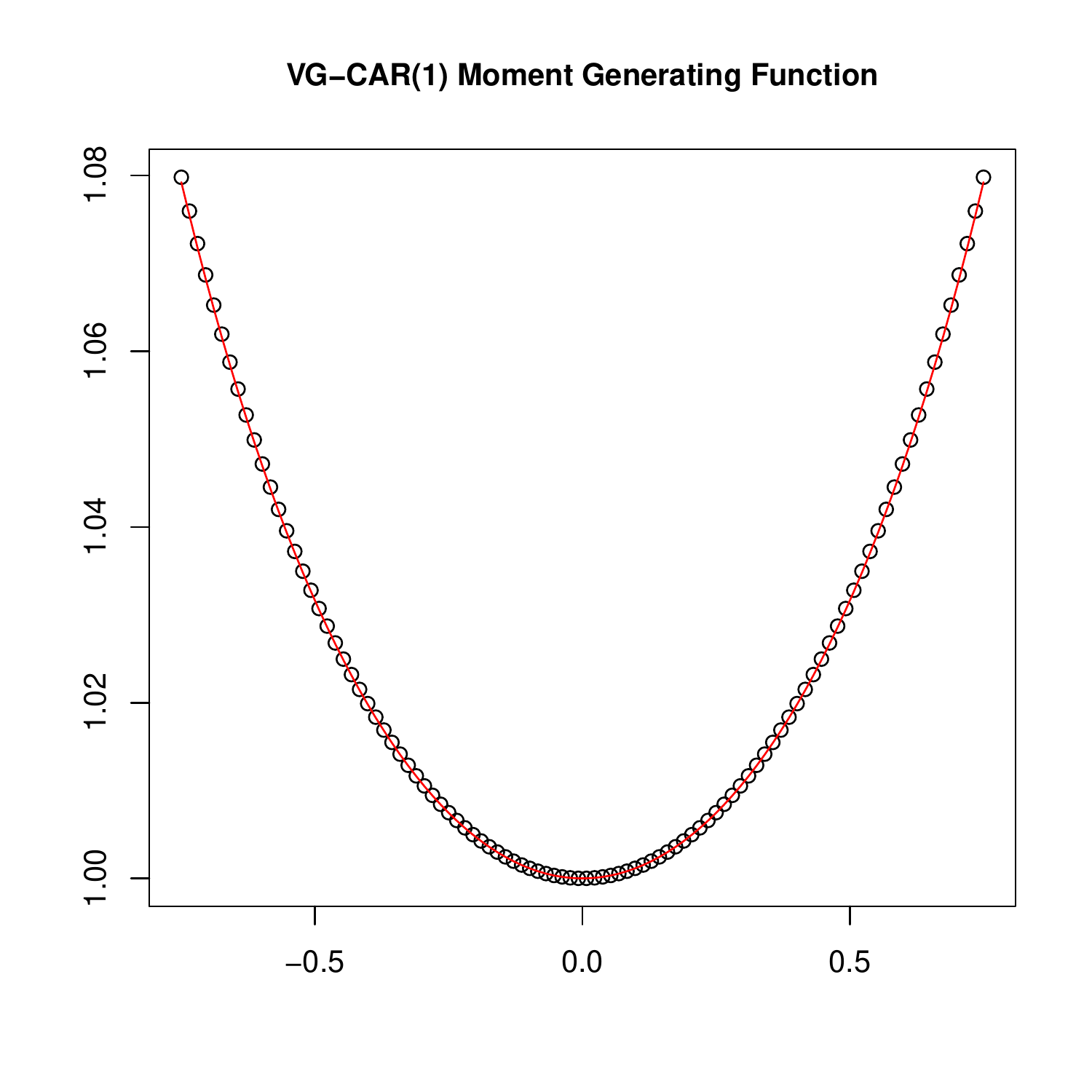}
\caption{\label{fig:Fig2} Comparison of theoretical and
approximated moment generating function for a VG-CAR(1) model}
\end{figure}

\noindent Let us define
\(V_{t_0}^{t}\) as: 
\begin{equation}
V_{t_0}^{t}=\int_{t_0}^{t}e^{-2a\left(t-u\right)}\mbox{d}\Lambda_{u}.
\label{eq:newV}
\end{equation} 
We can approximate the integral in \eqref{eq:newV} with a left Riemann sum as follows:
\begin{equation}
V_{t_0}^{t}\approx V_{t_0}^{t}\left(n\right)= \underset{k=0}{\stackrel{\left[2^n\left(t-t_0\right)\right]-1}{\sum}}e^{-2a\left(t-t_0-k2^{-n}\right)}\left(\Lambda_{t_0+\left(k+1\right)2^{-n}}-\Lambda_{t_0+k2^{-n}}\right).
\label{eq:ApproxV}
\end{equation} The increments
\(\Lambda_{t_0+\left(k+1\right)2^{-n}}-\Lambda_{t_0+k2^{-n}}\) in \eqref{eq:ApproxV}
have a density of the shape in \eqref{2}. Therefore we can approximate
these densities using the Laguerre polynomials. To this aim, we first introduce a discrete random variable
\(\mathcal{U}_k\):\[
\mathcal{U}_k=\left\{
\begin{array}{lll}
u_1 & & \mathbb{P}\left(u_1\right)\\
\vdots & & \vdots\\
u_m & & \mathbb{P}\left(u_m\right)
\end{array}
\right. 
\]that approximates the $k-th$ increment
\(\Lambda_{t_0+\left(k+1\right)2^{-n}}-\Lambda_{t_0+k2^{-n}}\).  The random variable \(V_{t_0}^{t}\left(n\right)\) can be approximated
introducing the new random variable \(V_{t_0}^{t}\left(n,m\right)\)
defined using dyadic Riemann sums reads: 
\begin{equation}
V_{t_0}^{t}\left(n,m\right)=\left\{
\begin{array}{lll}
\underset{k=0}{\stackrel{\left[2^n\left(t-t_0\right)\right]-1}{\sum}} e^{-2a\left(t-t_0-k2^{-n}\right)}u_1 & \left[2^n\left(t-t_0\right)\right]-1,0,\ldots,0 & \mathbb{P}^{\left[2^n\left(t-t_0\right)\right]-1}\left(u_1\right)\\
\vdots & & \vdots\\
\underset{k=0}{\stackrel{\left[2^n\left(t-t_0\right)\right]-1}{\sum}} e^{-2a\left(t-t_0-k2^{-n}\right)}u_k & n_1,\ldots,n_m &  \underset{i=1}{\stackrel{m}{\prod}}\mathbb{P}^{n_i}\left(u_i\right)\\
\vdots & & \vdots\\
\underset{k=0}{\stackrel{\left[2^n\left(t-t_0\right)\right]-1}{\sum}} e^{-2a\left(t-t_0-k2^{-n}\right)}u_m & 0,\ldots,0,\left[2^n\left(t-t_0\right)\right]-1 & \mathbb{P}^{\left[2^n\left(t-t_0\right)\right]-1}\left(u_m\right)
\end{array}
\right. .
\label{ApproxVOUTCBmAP}
\end{equation} 
Observe that the random variable $V_{t_0}^{t}\left(n,m\right)$ has \(m^{\left[2^n\left(t-t_0\right)\right]-1}\)
realizations. Denoting \(V_{t_0}^{t}\left(n,m,i\right)\) the i\(-th\)
realization of the random variable \(V_{t_0}^{t}\left(n,m\right)\) and
\(\mathbb{P}\left[V_{t_0}^{t}\left(n,m,i\right)\right]\) its
probability, we obtain the following approximated density:
\begin{equation}
f_{Y_t\left|\mathcal{F}_{t_0}\right.}\left(y\right)=\sum_{i=1}^{m^{\left[2^n\left(t-t_0\right)\right]-1}} \phi\left(y,y_0e^{-a\left(t-t_0\right)},V_{t_0}^{t}\left(n,m,i\right)\right)\mathbb{P}\left[V_{t_0}^{t}\left(n,m,i\right)\right].
\label{eq:ApproxDens}
\end{equation} To check the accuracy of this approximation, we compare
the theoretical moment generating function of an Ornstein-Uhlenbeck
driven by a Variance Gamma model obtained through the result in \cite{Hitaj2019} with the moment generating function of the finite mixture of normals with density \eqref{eq:ApproxDens}.\newline
Figure \ref{fig:Fig2} reports a graphical comparison of the theoretical
and the approximated moment generating function of a VG-CAR(1) with
\(a = 0.25\), \(t=\frac14\) and \(t_0\). The interval
\(\left[t_0,t\right)\) has been divided into subintervals of length
\(\Delta t =2^{-6}\approx0.01562\) and fixing \(m=2\) we get 65536
realizations of the random variable \(V_{t_0}^{t}\left(n,m\right)\).

The result in \eqref{eq:ApproxDens} can be used to construct a Maximum
Likelihood Estimation procedure. In the following we perform a simulation and estimation study for the VG-CAR(1) model. As
benchmark we use the Quasi-Gaussian Likelihood method extended to the
SDE driven by a standardized L\'evy noise introduced in \cite{Masuda2013}. We
perform the following steps:

\begin{enumerate}
\item We simulate a sample for a VG-CAR(1) model where $a=0.25$ while the distribution at time 1 of the subordinator process is $\Gamma\left(1,1\right)$. In the simulation we use the Euler-Maruyama method with a frequency $\Delta t = 0.01$.
\item We get a new trajectory by subsampling the data obtained at the previous point with a lower frequency, i.e. $\Delta t = 1$.
\item We estimate the parameters, using the data obtained in step 2, by maximizing the log-likelihood constructed using the Laguerre approximation.
\end{enumerate}

\begin{figure}[htbp]
	\centering
		\includegraphics[width=0.50\textwidth]{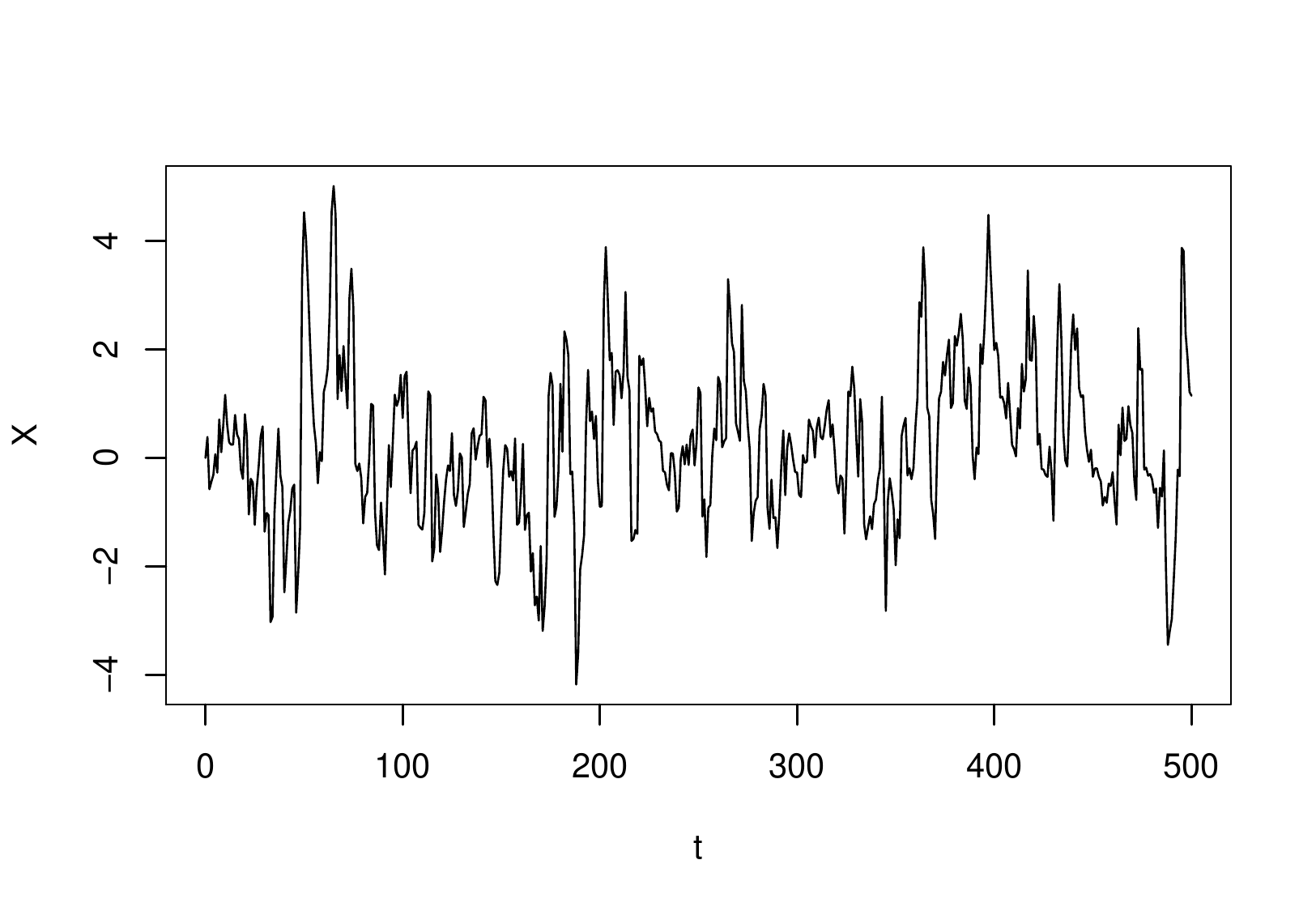}
		\caption{Sample path of a VG-OU process.\label{Fig:SampleOU}}
\end{figure}

\begin{verbatim}
##         b         a 		 Shape 
## 0.2226184 0.9900000 --------- # YUIMA ESTIMATION QMLE BASED ON MASUDA
## 0.2394667 1.0822139 1.0501550 # Estimation Based on Gauss Laguerre Quadrature
## 0.2400000 1.0000000 1.0000000 # TRUE PARAMETERS
\end{verbatim}

\subsection{Estimation of a Gaussian CARMA(p,q) model.}\label{estimation-of-a-gaussian-carmapq-model.}

In this section we review the literature for the estimation methods of
CARMA(p,q) model driven by a Brownian Motion. As discussed in \cite{Tomasson2015}, we have two different approaches for the estimation of a Gaussian CARMA process. The first is based on the frequency
domain representation of the CARMA process. The estimated
parameters are obtained by minimizing a distance between theoretical
\(f\left(\omega\right)\) and empirical \(\hat{f}\left(\omega\right)\)
spectral density, for instance: \[
\underset{a_1,\ldots, a_p \ \ \ b_1,\ldots,b_q}{\text{argmin}}\int_{-\infty}^{+\infty}\left\{\ln\left[f\left(\omega\right)\right]+\frac{\hat{f}\left(\omega\right)}{f\left(\omega\right)}\right\}\mbox{d}\omega 
\] where \[
f\left(\omega\right)=\frac{\mathbf{b}\left(i\omega\right)\mathbf{b}\left(-i\omega\right)}{2\pi\mathbf{a}\left(i\omega\right)\mathbf{b}\left(-i\omega\right)}.
\] The alternative estimation approach is based on the time domain
representation of the CARMA process. In this case, the unobservable
state process can be extrapolated using the Kalman filter therefore we
get the estimates for the model parameters by maximizing the
loglikelihood function or minimizing the least-squares error. A detailed
description of the Kalman filter and the construction of the gaussian
loglikelihood function can be found in \cite{Iacus2015}.

\subsection{Estimation of a L\'evy CARMA(p,q) model driven by a Time Changed Brownian Motion.}\label{estimation-of-a-luxe9vy-carmapq-model-driven-by-a-time-changed-brownian-motion.}

Here we discuss how to estimate the CARMA(p,q) model when the driving
noise is a Time Changed Brownian Motion. In this case we propose two
alternatives. The first approach combines the Kalman Filter with the
approximation transition density of the CARMA(p,q) process while the
second use the methodology for recovering noise with the estimation
method discussed for the Normal Variance Mean Mixture.

\subsubsection{L\'evy CARMA estimation using the approximated transition density}\label{luxe9vy-carma-estimation-using-the-approximated-transition-density}

In order to obtain an approximated transition density for a
CARMA(p,\ q) process we first need to determine the conditional mean and
the conditional variance of the state process \(X_t\) given the information
contained in the \(\sigma\)-field \(\mathcal{G}_{t_0,t}\) and  the
state process at \(X_{t_0}\) defined respectively as: \[
\mathbb{E}\left[X_t\left|\mathcal{G}_{t_0,t},X_{t_0}\right.\right]=e^{\mathbf{A}\left(t-t_0\right)}X_{t_0}.
\] \[
\mathbb{V}\text{ar}\left[X_t\left|\mathcal{G}_{t_0,t},X_{t_0}\right.\right]=\int_{t_0}^{t}e^{\mathbf{A}\left(t-u\right)}\mathbf{e}\mathbf{e}^{\top}e^{\mathbf{A}^\top\left(t-u\right)}\mbox{d}\Lambda_u.
\] Therefore the transition density of the CARMA(p,q) model \(Y_t\)
given \(\mathcal{G}_{t_0,t}\) and \(X_{t_0}\) is: \[
Y_t\left|\left(\mathcal{G}_{t_0,t},X_{t_0}\right.\right)\sim N\left(\mathbf{b}^\top e^{\mathbf{A}\left(t-t_0\right)}X_{t_0}, \int_{t_0}^{t}\mathbf{b}e^{\mathbf{A}\left(t-u\right)}\mathbf{e}\mathbf{e}^{\top}e^{\mathbf{A}^\top\left(t-u\right)}\mathbf{b}^\top \mbox{d}\Lambda_u \right)
\] Defining the quantity
\(V_{t_0}^{t}=\int_{t_0}^{t}\mathbf{b}e^{\mathbf{A}\left(t-u\right)}\mathbf{e}\mathbf{e}^{\top}e^{\mathbf{A}^\top\left(t-u\right)}\mathbf{b}^\top \mbox{d}\Lambda_u\),
the transition density of the CARMA(p,q) process \(Y_t\) given
\(X_{t_0}\) can be written in the following form: \begin{equation}
f_{Y_t\left|X_{t_0}\right.}\left(y\right)=\int_{0}^{+\infty} \varphi\left(y;\mathbf{b}e^{\mathbf{A}\left(t-t_0\right)}X_{t_0},v\right)g_{V_{t_0}^{t}}\left(v\right)\mbox{d}v,
\label{eq:RealDensityCarma}
\end{equation} where \(\varphi\left(y,\mu,\sigma^2\right)\) is a normal
density with mean \(\mu\) and variance \(\sigma^2\);
\(g_{V_{t_0}^t}\left(v\right)\) is the density of \(V_{t_0}^{t}\). As
done in Section \ref{Est:OUdrivenTCBM}, we approximate the integral in
\(V_{t_0}^t\) with a left Reimann sum and we have: \begin{equation}
V_{t_0}^t\approx V_{t_0}^{t}\left(n,m\right)=\left\{
\begin{array}{lll}
\underset{k=0}{\stackrel{\left[2^n\left(t-t_0\right)\right]-1}{\sum}}
\mathbf{b}e^{\mathbf{A}\left(t-t_0-k2^{-n}\right)}\mathbf{e}\mathbf{e}^{\top}e^{\mathbf{A}^\top\left(t-t_0-k2^{-n}\right)}\mathbf{b}^\top
u_1 & \left[2^n\left(t-t_0\right)\right]-1,0,\ldots,0 & \mathbb{P}^{\left[2^n\left(t-t_0\right)\right]-1}\left(u_1\right)\\
\vdots & &\vdots\\
\underset{k=0}{\stackrel{\left[2^n\left(t-t_0\right)\right]-1}{\sum}} \mathbf{b}e^{\mathbf{A}\left(t-t_0-k2^{-n}\right)}\mathbf{e}\mathbf{e}^{\top}e^{\mathbf{A}^\top\left(t-t_0-k2^{-n}\right)}\mathbf{b}^\top u_k & n_1,\ldots,n_m &  \underset{i=1}{\stackrel{m}{\prod}}\mathbb{P}^{n_i}\left(u_i\right)\\
\vdots & & \vdots\\
\underset{k=0}{\stackrel{\left[2^n\left(t-t_0\right)\right]-1}{\sum}} \mathbf{b}e^{\mathbf{A}\left(t-t_0-k2^{-n}\right)}\mathbf{e}\mathbf{e}^{\top}e^{\mathbf{A}^\top\left(t-t_0-k2^{-n}\right)}\mathbf{b}^\top u_m & 0,\ldots,0,\left[2^n\left(t-t_0\right)\right]-1 & \mathbb{P}^{\left[2^n\left(t-t_0\right)\right]-1}\left(u_m\right)
\end{array}
\right. ,
\label{eq:ApproxVarCar}
\end{equation} Thus \(f_{Y_t\left|X_{t_0}\right.}\left(y\right)\)
can be approximated with the finite mixture density function
\(\hat{f}_{Y_{t}\left|X_{t_0}\right.}\left(y\right)\) that reads: \begin{equation}
\hat{f}_{Y_{t}\left|X_{t_0}\right.}\left(y\right)=\sum_{i=1}^{m^{\left[2^n\left(t-t_0\right)\right]-1}} \phi\left(y,\mathbf{b}^\top e^{\mathbf{A}\left(t-t_0\right)}X_{t_0},V_{t_0}^{t}\left(n,m,i\right)\right)\mathbb{P}\left[V_{t_0}^{t}\left(n,m,i\right)\right],
\label{eq:ApproxDensCarma}
\end{equation} where \(V_{t_0}^t\left(n,m,i\right)\) denotes the
\(i-th\) realization of the random variable
\(V_{t_0}^{t}\left(n,m\right)\) in \eqref{eq:ApproxVarCar}.\newline For the approximated loglikelihood fuction
\(\hat{\mathcal{L}}\left(\theta\right)\) we need to infer the state
process \(X_t\). From the estimated process \(\hat{X}_t\),
we can determine the optimal value for the parameter vector \(\theta\) solving the
following optimization problem

\begin{equation*}
\theta=\text{argmax} \sum_{i=1}^{N}\ln\left[\hat{f}_{Y_{t_i}\left|\hat{X}_{t_{i-1}}\right.}\left(y_{t_i}\right)\right].
\end{equation*} In this paper we consider two alternatives for the estimation of the
state process \(X_t\): the Kalman Filter and the filtering approach
discussed in Section \ref{LCARMA} and proposed in \cite{Brockwell2011}. 


In the following table we compare the GQMLE approach discussed in \cite{Iacus2015} for a General L\'evy CARMA(p,q) model
and our approaches. The labels \texttt{GL-HF} and \texttt{GL-HFKF} denote the Maximum Likelihood estimation method based on our approximated transition density, the only difference is related to the method for filtering the state process from the observable data. In \texttt{GL-HF} case, the estimated state process $\left\{\hat{X}_{t}\right\}_{t\geq0}$ is obtained using the dynamic in \eqref{eq:StochFilterBrock} [see \cite{Brockwell2011} for more information] while in \texttt{GL-HFKF} case the standard Kalman Filter is used.
\begin{verbatim}
##         a1         a2         b0         b1      Shape      Scale  
## 1.35000000 0.05000000 0.20000000 1.00000000 1.00000000 1.00000000 # True Parameters
## 1.38164866 0.04634073 0.18808589 0.99993332 1.15596369 1.00265283 # GL-HF 
## 1.31162953 0.04494326 0.19126241 0.98709469 1.12985742 1.01003225 # GL-HFKF 
## 1.35175679 0.04813515 0.18653096 1.03154645 ---------- ---------- # GQMLE
\end{verbatim}

\section{Option Pricing in a L\'evy CARMA(p,q) model.}\label{option-pricing-in-a-luxe9vy-carmapq-model.}

In this section we discuss, using the approximated transition density,
how to evaluate the expected value of the transformation
\(g\left(X_T\right)\) where \(X_T\) can be a Normal Variance Mean Mixture or a CARMA with a Time Changed Brownian Motion driving noise.

In the Normal Variance Mean Mixture case we discuss also the behaviour of the
error term while in the second case we analyze it by a comparison with
the Monte Carlo simulation. The result here can be applied to extend the
option pricing formula for options on futures contracts proposed in
\cite{PASCHKE20102742} for the gaussian CARMA model. This approach can be used also
for the evaluation of the term structure of futures.

\subsection{Normal Variance Mean Mixture} \label{OptNVMM}

Starting from the formal definition of Normal Variance Mean Mixture in
\eqref{1}, we define the sequence of function
\(\mathsf{E}\left[g\left(X_T^{m}\right)\left|\mathcal{F}_{0}\right.\right]\)
as following: \begin{equation}
\sum_{i=1}^{m}\mathsf{E}\left[g\left(\mu+\theta \Lambda_m+\sqrt{\Lambda_m}Z\right)\left|\mathcal{F}_0,\Lambda_m=u_i\right.\right]\mathbb{P}\left(u_i\right)
\label{eq:SectN1}
\end{equation} where \(\Lambda_m\) and \(\mathbb{P}\left(u_i\right)\) are
defined in \eqref{eq:approxMixRVNVMM}. The quantity
\(\left[g\left(\mu+\theta \Lambda_n+\sqrt{\Lambda_n}Z\right)\left|\mathcal{F}_0,\Lambda_n=u_i\right.\right]\)
is the expectation of a gaussian distribution with mean
\(\mu+\theta \Lambda_n\) and variance \(\Lambda_n\). 

The formulas proposed in this section can be applied for the evaluation
of the contingent claim when the underlying is a transformation of a
Time Change Brownian Motion. In the next section we show a comparison of
our approach with a Monte Carlo simulation when the log price is a
Variance Gamma process and the function \(g\) is the final payoff of a
European Call Option.

\subsubsection{Simulation Comparison}\label{simulation-comparison}

\begin{figure}[!htbp]
\centering
\includegraphics[width=0.5\textwidth]{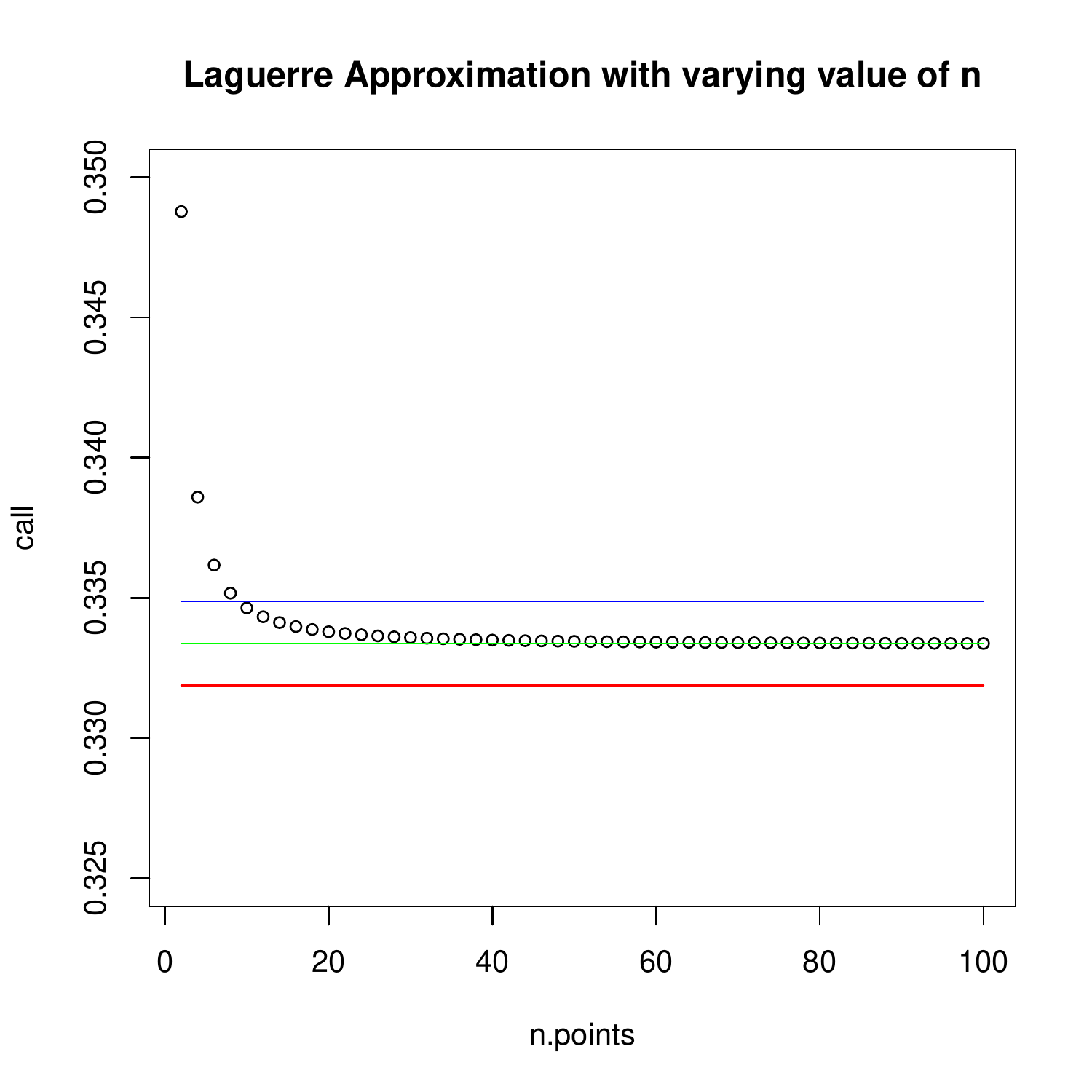}
\caption{Comparison of prices obtained using Monte Carlo
simulation and the Laguerre Option pricing formula \eqref{eq:SectN1} for an ATM
European Call option. \label{fig:FigOptionPriVar1}}
\end{figure}

Figure \ref{fig:FigOptionPriVar1} shows the behaviour of a European Call
option price for varying value of \(n\) in the Gauss-Laguerre
approximation approach. In this example the model parameters are \(r=0\),
\(\theta=-0.5\), \(\alpha=1\), \(\beta=1\), underlying price \(S_0=1\)
and time to maturity \(T=1\).

\begin{figure}[!htbp]
\centering
\includegraphics[width=0.5\textwidth]{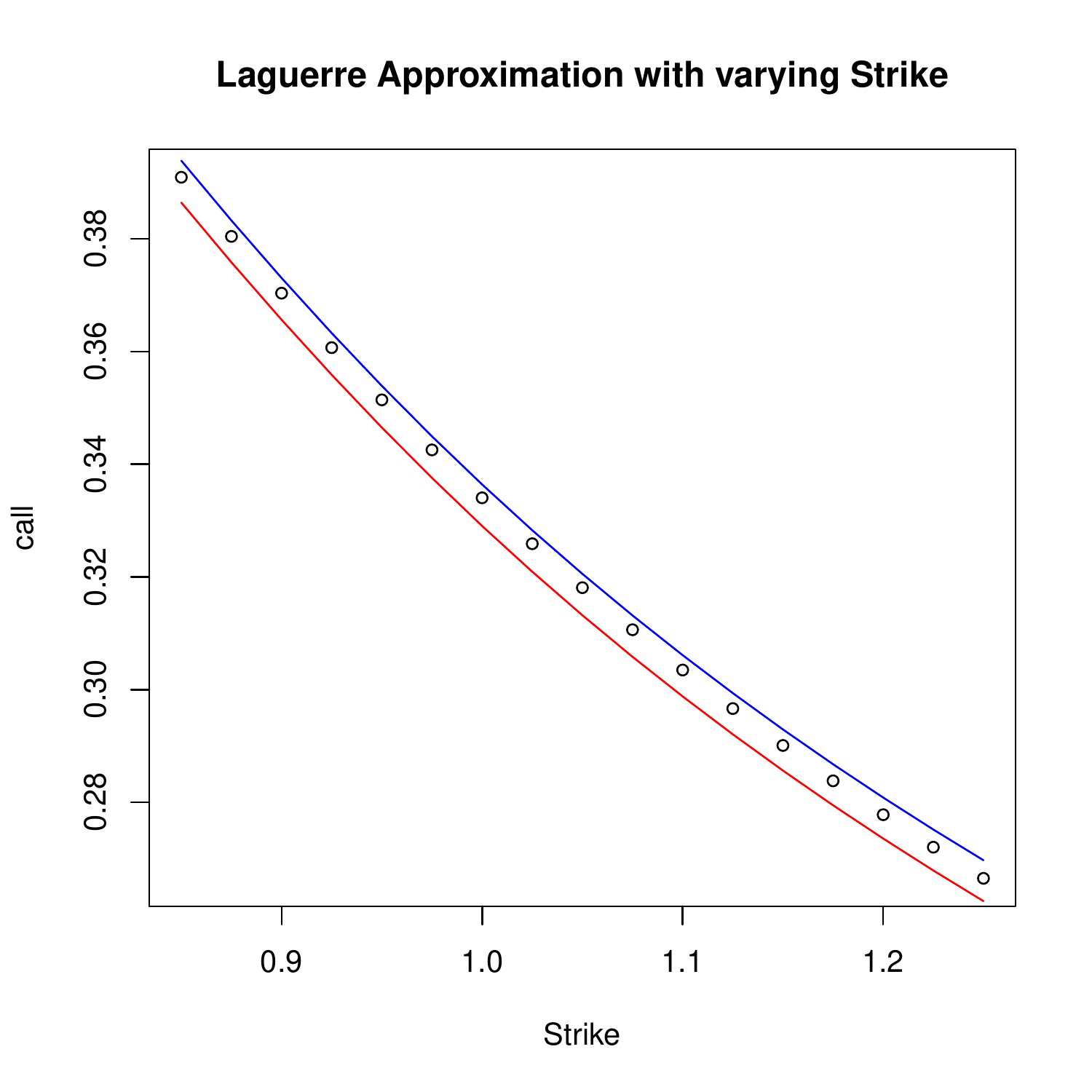}
\caption{Comparison of price obtained using Monte Carlo
simulation and the Laguerre Option pricing formula \eqref{eq:SectN1} for different levels
of strike price.\label{fig:FigOptionPriVar2}}
\end{figure}

\begin{figure}[!htbp]
\centering
\includegraphics[width=0.5\textwidth]{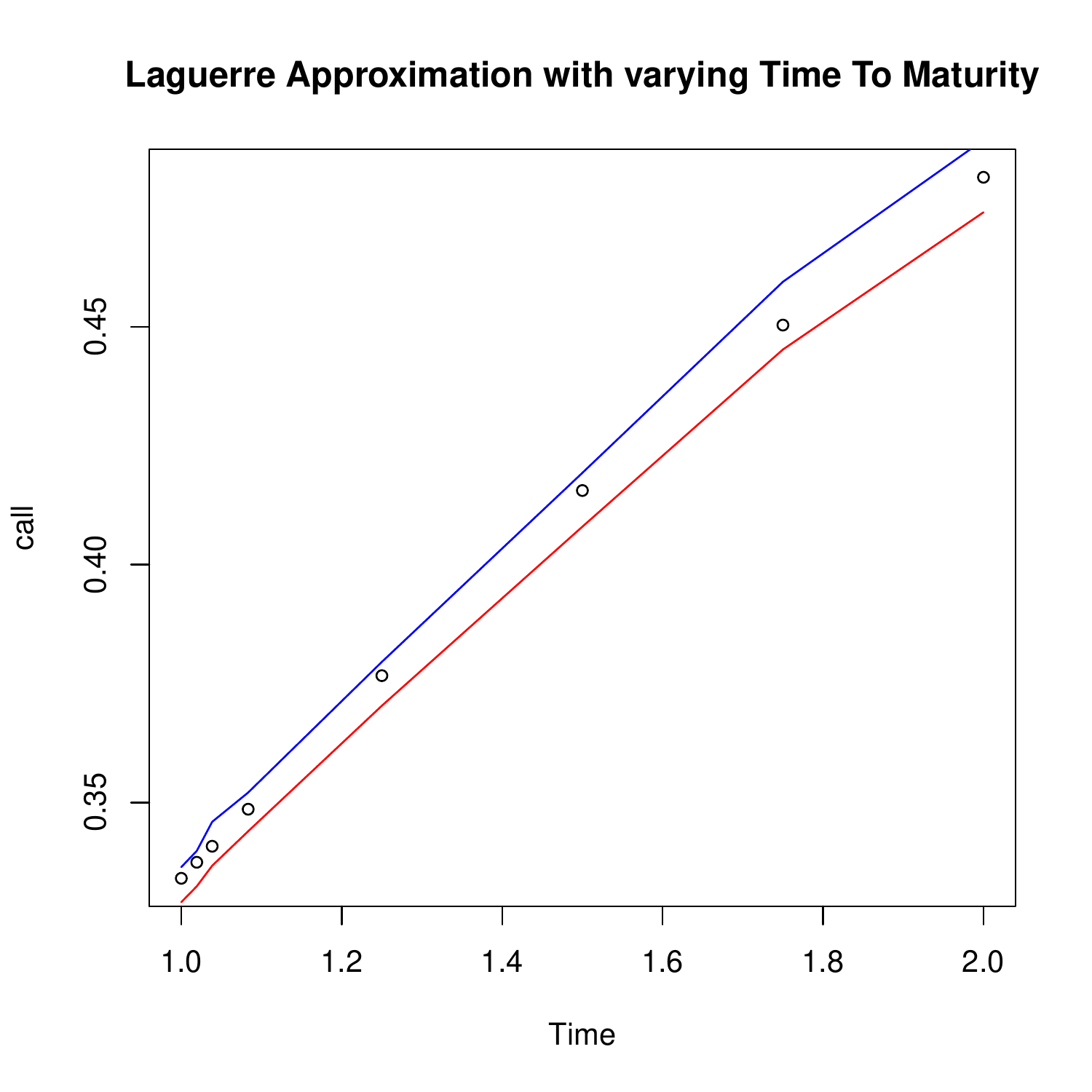}
\caption{Comparison of prices obtained using Monte Carlo
simulation and the Laguerre Option pricing formula \eqref{eq:SectN1} for different levels
of strike price.\label{fig:FigOptionPriVar3} }
\end{figure}

We analyze also the behaviour of the approximation for different 
strike levels in Figure \ref{fig:FigOptionPriVar2} and for varying Time to
maturity in Figure \ref{fig:FigOptionPriVar3}. In the latter it is
important to satisfy the condition \(\alpha T\geq1\) otherwise we need
to use the Generalized Gauss Laguerre approximation due to the presence
of a no negligible singularity in the Mixing Gamma random Variable at
point zero.

\clearpage
\subsection{Time Changed CARMA process}\label{time-changed-carma-process}

We discuss here how to extend the general result in Section
\ref{OptNVMM} for the Time Changed Brownian Motion to the TCBm-CARMA
process. The main idea is to use the approximation of $V^t_{t_0}$ introduced in Equation \eqref{eq:RealDensityCarma}. 
The general pricing formula of the final payoff \(g\left(Y_T\right)\)
can be derived following the same steps as in the previous section. The resulting formula reads:
\begin{equation}
\mathsf{E}\left[g\left(Y_{T}\right)\left|\mathcal{F}_{t_0}\right.\right]=\sum_{k=1}^{m^{\left[2^n\left(T-t_0\right)\right]-1}}\mathsf{E}\left[g\left(Y_{T}^{m,n}\right)\left|\mathcal{F}_{t_0},V_{t_0}^{T}=V_{t_0}^{T}\left(m,n,k\right)\right.\right]\mathbb{P}\left(V_{t_0}^{T}\left(m,n,k\right)\right),
\label{mainResCarmaPricing}
\end{equation} where
\(Y_{T}^{m,n}\left|\mathcal{F}_{t_0},V_{t_0}^{T}=V_{t_0}^{T}\left(m,n,k\right)\right.\sim N\left(b^{\top}e^{\mathbf{A}\left(T-t_0\right)}X_{t_0},V_{t_0}^{T}\left(m,n\right)\right)\).

This result can easily find applications in different financial modeling topics such as the construction of futures term
structure, option pricing of bond pricing under the hypothesis that the dynamics of the 
underlying follows a Time Change CARMA model.

\subsubsection{Futures Term Structure with a TCBm CARMA(p,q) model}\label{futures-term-structure-with-a-tcbm-carmapq-model}

In the filtered probability space we assume that it exists an equivalent martingale measure $\mathbb{Q}\sim\mathbb{P}$ exists. We also assume that the price $S_t$ of the commodity asset follows an exponential TCBm-CARMA(p,q) model under the measure $\mathbb{Q}$ defined as:  
\[
S_t = S_{t_0} e^{Y_{t}},
\]
where $Y_t$ is a CARMA(p,q) model described in Section \ref{LCARMA}; the driving noise in a Time Change Brownian motion i.e.
\[
L_t=W_{\Lambda_t}
\]
where $W_t$ is a Brownian Motion and $\Lambda_t$ is an independent subordinator process with cumulant generating function $k_\Lambda\left(u\right)$ defined as:
\[
k_\Lambda\left(u\right):=\ln\left[\mathsf{E}\left(e^{u \Lambda_1}\right)\right].
\]
Arbitrage theory is based on the assumption that price of a future should be equal to the expected value of the  price at maturity under the risk neutral measure $\mathbb{Q}$. Therefore, the log future price with maturity $T\geq t_0$ can be written as:
\begin{equation}\label{q-measure}
\ln F^T_{t_0} = \ln \mathsf{E}^{\mathbb{Q}}\left[ S_{T}\left|\mathcal{F}_{t_0}\right.\right] 
\end{equation}
Defining the $\sigma$-field $\mathcal{G}^{t}_{t_0}=\sigma\left(\mathcal{F}_{t_0} \cup \sigma\left(\left\{\Lambda_u\right\}_{u\leq t}\right)\right)$ with $t\geq t_0$ we have:
\[
W_{\Lambda_t}-W_{\Lambda_{t_0}}\left|\mathcal{G}^t_{t_0}\right.\sim N\left(0,\Lambda_t-\Lambda_{t_0}\right).
\]
Using the iterative property of the conditional expected value, equation \eqref{q-measure} can be rewritten as:
\begin{equation}\label{q-measure exp}
\ln F^T_{t_0} = \ln \mathsf{E}^{\mathbb{Q}}\left[\mathsf{E}^{\mathbb{Q}}\left( S_{T}\right|\mathcal{G}^{T}_{t_0})|\mathcal{F}_{t_0}\right]. 
\end{equation}
It is worth to notice that the random variable $\ln S_{T}\left|\mathcal{G}^{T}_{t_0}\right.$ is normally distributed. Therefore, we have that: 
\[
\mathsf{E}^\mathbb{Q}\left(S_{T}\right|\mathcal{G}^{T}_{t_0}) = \exp\left(\ln S_{t_{0}} + \mathsf{E}^{\mathbb{Q}}\left[\ln S_{T}|\mathcal{G}^{T}_{t_0}\right] + \frac{1}{2} \mathsf{VAR}^\mathbb{Q}\left[\ln S_{T}|\mathcal{G}^{T}_{t_0}\right]\right).
\]
Then:
\begin{equation}\label{second step}
\ln F^T_{t_0} = \ln \mathsf{E}^{\mathbb{Q}}_{t}\left[ e^{\ln S_{t_{0}}} e^{\mathsf{E}^\mathbb{Q}\left[\ln S_{T}|\mathcal{G}^{T}_{t_0}\right] + \frac{1}{2} \mathsf{VAR}^\mathbb{Q}\left[\ln S_{T}|\mathcal{G}^{T}_{t_0}\right]}|\mathcal{F}_{t_0}\right],
\end{equation}
 and rearranging:
\begin{equation}\label{third step}
\ln F^T_{t_0} = \ln S_{t_{0}} + \ln \left[\mathsf{E}^{\mathbb{Q}}\left(e^{\mathsf{E}^\mathbb{Q}\left[\ln S_{T}|\mathcal{G}^{T}_{t_0}\right] + \frac{1}{2} \mathsf{VAR}^\mathbb{Q}\left[\ln S_{T}|\mathcal{G}^{T}_{t_0}\right]}\right)|\mathcal{F}_{t_0}\right].
\end{equation}
At this stage, it is possible to introduce the conditional transition density of a CARMA(p,q) model driven by a Time Changed Brownian Motion $Y_T$ given $\mathcal{G}^{T}_{t_0}$ as:
\[
Y_T\left| \mathcal{G}^T_{t_0}\right.\sim N\left(\mathbf{b}^\top e^{\mathbf{A}\left(T-t_0\right)}X_{t_0}, \int_{t_0}^{T}\mathbf{b}^{\top}e^{\mathbf{A}\left(T-u\right)}\mathbf{e}\mathbf{e}^{\top}e^{\mathbf{A}^\top\left(T-u\right)}\mathbf{b} \mbox{d}\Lambda_u \right)
\]
Given this result, we obtain:
\begin{equation}\label{fourth step}
\ln F^{T}_{t_0}= \ln S_{t_{0}} + \ln \left[\mathsf{E}^{\mathbb{Q}}\left(e^{\mathbf{b}^\top e^{\mathbf{A}\left(T-t_0\right)}X_{t_{0}} + \frac{1}{2} \int_{t_0}^{T}\mathbf{b}^{\top}e^{\mathbf{A}\left(T-u\right)}\mathbf{e}\mathbf{e}^{\top}e^{\mathbf{A}^\top\left(T-u\right)}\mathbf{b} \mbox{d}\Lambda_u}|\mathcal{F}_{t_0}\right)\right].
\end{equation}
Simplifying:
\begin{equation}\label{fifth step}
\ln F^T_{t_0} = \ln S_{t_{0}} + \mathbf{b}^\top e^{\mathbf{A}\left(T-t_0\right)}X_{t_0} + \ln \left[\mathsf{E}^{\mathbb{Q}}\left(e^{\frac{1}{2} \int_{t_0}^{T}\mathbf{b}^{\top}e^{\mathbf{A}\left(T-u\right)}\mathbf{e}\mathbf{e}^{\top}e^{\mathbf{A}^\top\left(T-u\right)}\mathbf{b} \mbox{d}\Lambda_u}|\mathcal{F}_{t_0}\right)\right].
\end{equation}
We use the following theorem proposed in \cite{Eberlein1999} in order to evaluate the expected value in \eqref{fifth step}.
\begin{theorem}\label{EberlRaibleProp}
Let $\Lambda_{t}$ be a subordinator process with cumulant generating function
$k_\Lambda\left(u\right)$ and $f\left(u\right):\left[0,+\infty\right)\rightarrow\mathbb{C}$ be a complex left continuous function such that $\left|\mathsf{Re}\left(f\right)\right|\leq M$
then:
\begin{equation*}\label{EbRaiEq}
\mathsf{\mathsf{E}}\left[\exp\left(\int_{0}^{+\infty}f\left(u\right)\mbox{d}\Lambda_{u}\right)\right]=\exp\left(\int_{0}^{+\infty} k_{\Lambda}\left(f\left(u\right)\right)\mbox{d}u\right).
\end{equation*}
\end{theorem}
\noindent Using the above theorem and the following property of the cumulant function
\[
k_{\Lambda}\left(u \mathbbm{1}_{A}\right)=\mathbbm{1}_{A} k_{\Lambda}\left(u \right)
\]
we obtain the final result 
\begin{equation}
\ln F^{T}_{t_0} = \ln S_{t_{0}} + \mathbf{b}^\top e^{\mathbf{A}\left(T-t_0\right)}X_{t_0} +  \int_{t_0}^{T}  k_{\Lambda} \left(\frac{1}{2}\mathbf{b}^\top e^{\mathbf{A}\left(T-u\right)}\mathbf{e}\mathbf{e}^{\top}e^{\mathbf{A}^\top\left(T-u\right)}\mathbf{b} \right) \mbox{d}u .
\label{sixthstep}
\end{equation}

The approximated transition density of the TCBm-CARMA(p,q) model gives the possibility of evaluating the formulas in \eqref{sixthstep} in a easy way. By applying the general result in \eqref{mainResCarmaPricing} we get the following approximation:

\[
\ln F^{T}_{t_0}\left(m,n\right)=  \ln S_{t_{0}} + \mathbf{b}^\top e^{\mathbf{A}\left(T-t_0\right)}X_{t_0} +\ln \sum_{k=1}^{m^{\left[2^{n}\left(T-t_0\right)\right]-1}} e^{\frac12 V_{t_0}^{T}\left(m,n,k\right)}\mathbb{P}\left(V_{t_0}^{T}\left(m,n,k\right)\right)
\]

A numerical comparison of  the approximated formula
with the pricing results obtained through Monte Carlo simulation is reported below. The MC value is
evaluated using 10.000 simulated trajectories of a symmetric VG-CARMA(2,1)
model with autoregressive parameters \(a_1=1.4\) \(a_2=0.5\), moving
average parameters \(b_0=0.2\) \(b_1=1\) and  Gamma subordinator
process \(\left(\Lambda_{t}\right)_{t\geq0}\) with shape parameter
\(\alpha=1\) and scale parameter \(\beta=1\). The simulated trajectories
are obtained using the Euler discretization scheme for a L\'evy CARMA(p,q)
model as described in \cite{Iacus2015} on a regular grid with
\(\Delta t=\frac{T}{200}\) where \(T\) is the maturity of the
Future.\newline It is to worth to observe that since we have that \(\alpha T<1\), we can use the Generalized Gauss Laguerre Quadrature to avoid numerical issues that may arise due to the singularity at point 0.

\begin{figure}[!htbp]
\centering
\includegraphics[width=0.5\textwidth]{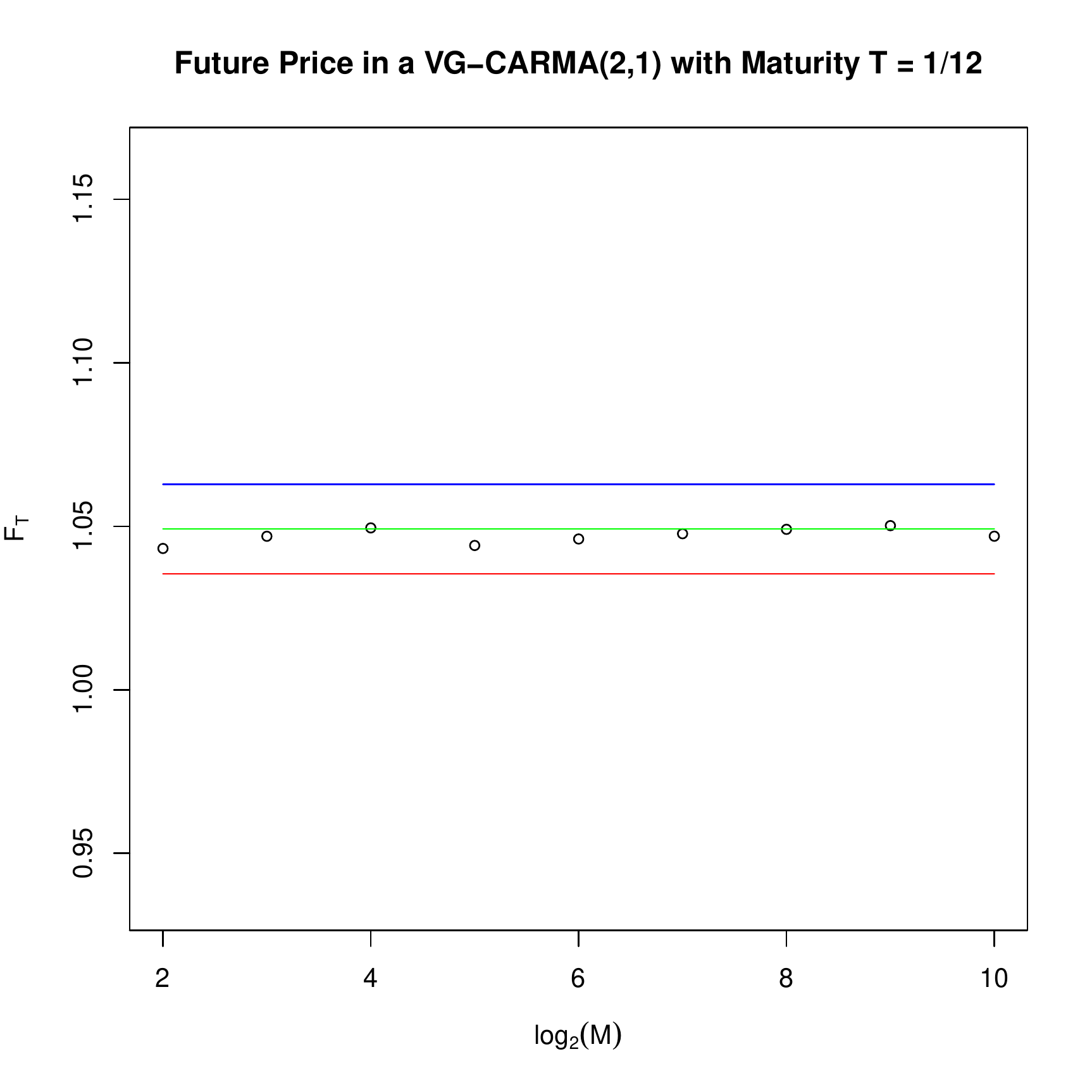}
\caption{Future price with maturity 1 month.\label{fut1m}}
\end{figure}

\begin{figure}[!htbp]
\centering
\includegraphics[width=0.5\textwidth]{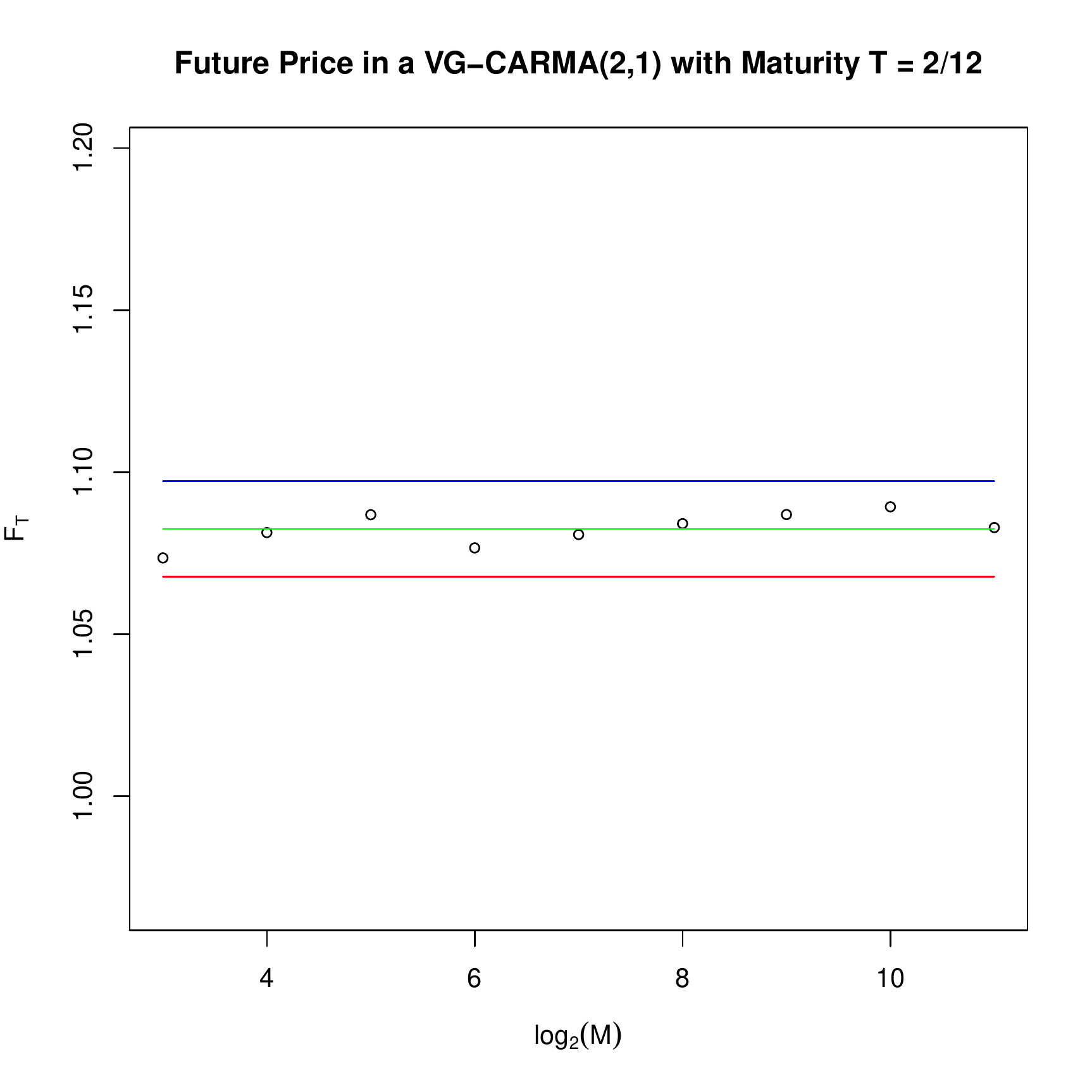}
\caption{Future price with maturity 2 months.\label{fut2m}}
\end{figure}

\begin{figure}[!htbp]
\centering
\includegraphics[width=0.5\textwidth]{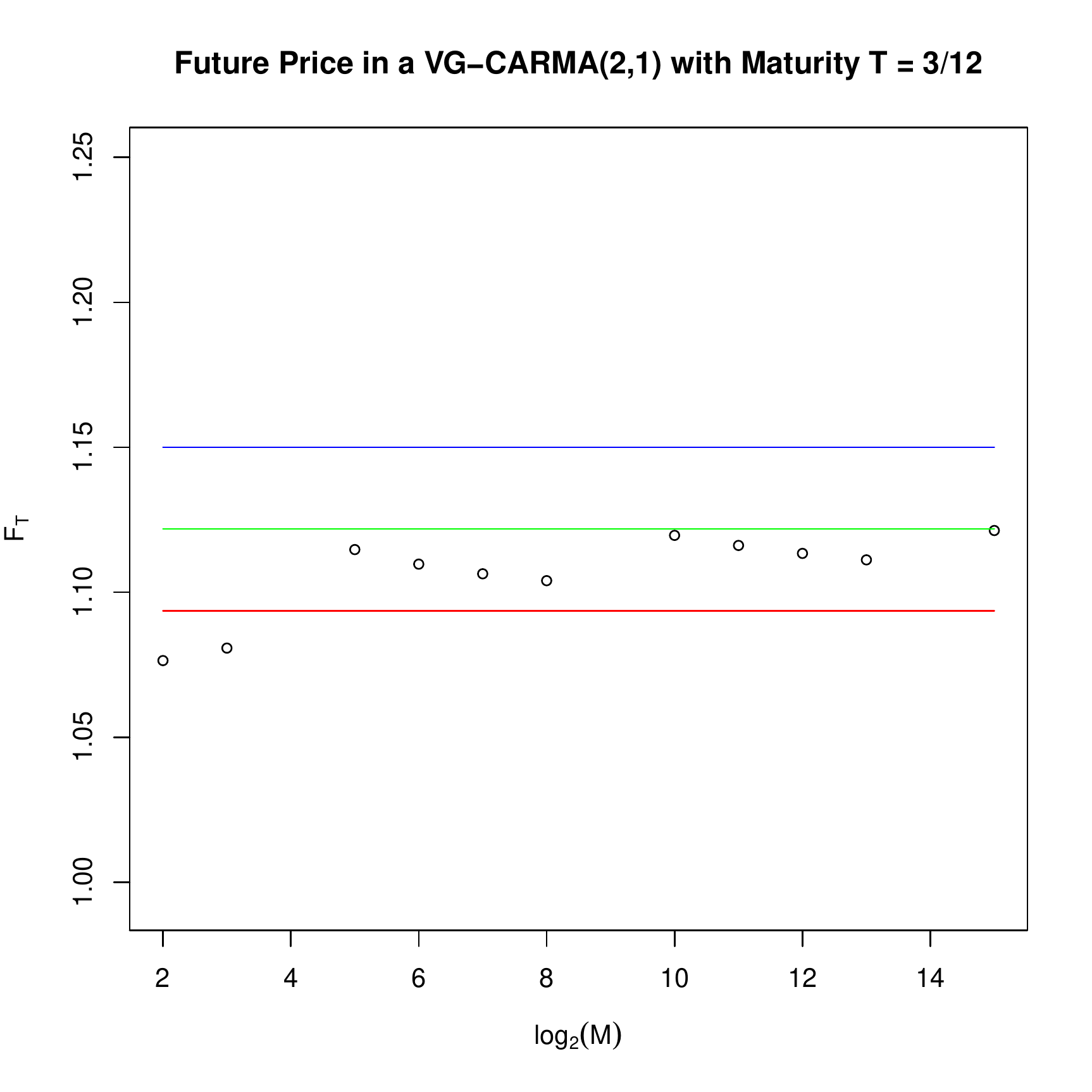}
\caption{Future price with maturity 3 months.\label{fut3m}}
\end{figure}

\begin{figure}[!htbp]
\centering
\includegraphics[width=0.5\textwidth]{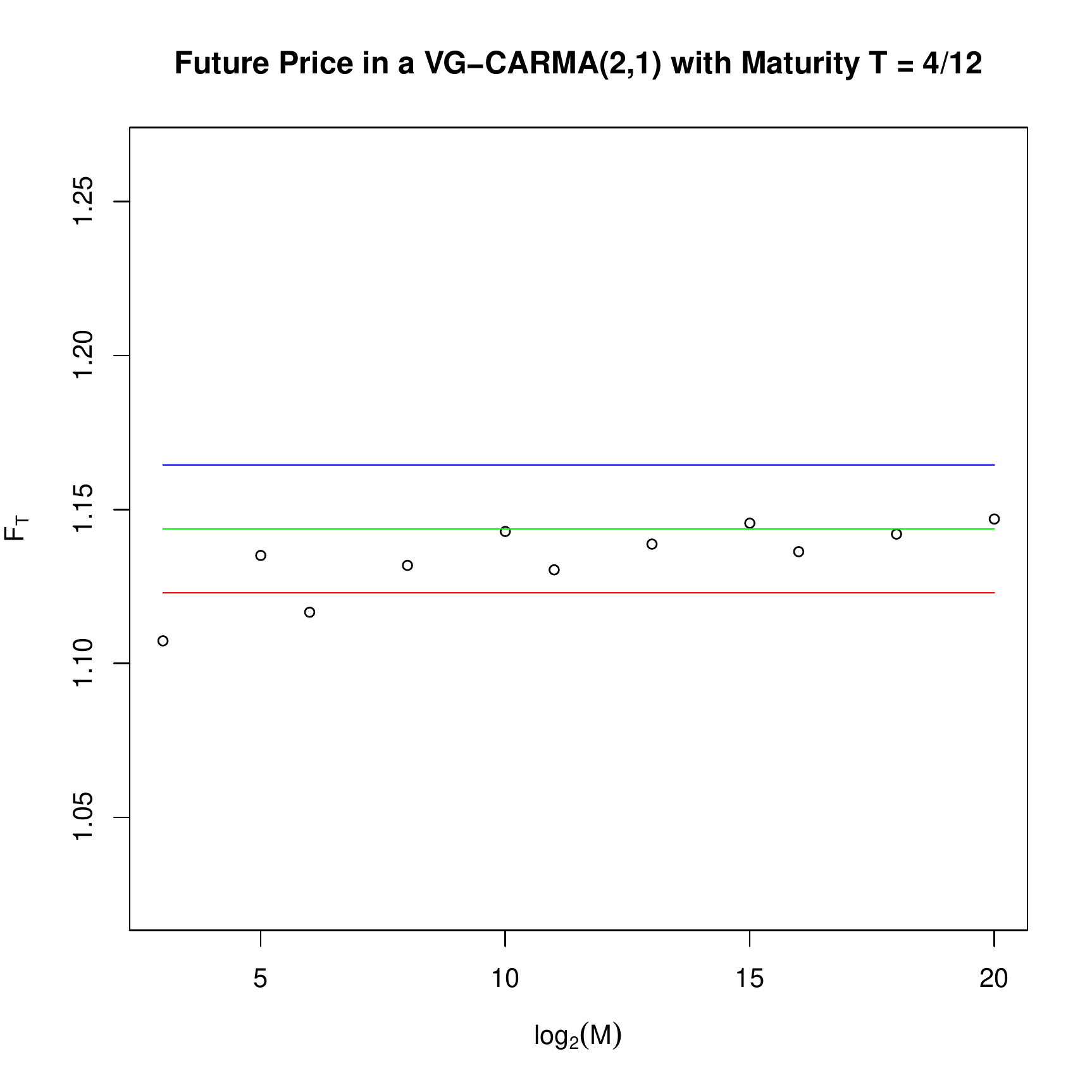}
\caption{Future Price with Maturity 4 Months.\label{fut4m}}
\end{figure}

See Table \ref{ComOptFuture} for the futures term structure and Figures \ref{fut1m}-\ref{fut4m} for an analysis based on the number of points $m$ used in the approximation.

\begin{table}
\centering
\begin{tabular}{lcccc}
\hline
T              & Lag.     & MC      & Ub      & Lb\\
\hline
\hline
$\frac{1}{12}$ & 1.04697  & 1.04918 & 1.06285 & 1.03550 \\
$\frac{2}{12}$ & 1.08293  & 1.08248 & 1.09719 & 1.06778 \\
$\frac{3}{12}$ & 1.12130  & 1.12183 & 1.15005 & 1.09361 \\
$\frac{4}{12}$ & 1.14691  & 1.14367 & 1.16442 & 1.12292 \\
\hline\hline\\
\end{tabular}
\caption{Pricing results for a future contract using MC and the approximated formula based on the Gauss-Laguerre quadrature.\label{ComOptFuture}}
\end{table}

\clearpage

\subsubsection{Futures Option Pricing formula in a TCBm CARMA(p,q) model}\label{futures-option-pricing-formula-in-a-tcbm-carmapq-model}

Here we discuss how to modify our general result in order to extend the
result about the Futures option prices in \cite{PASCHKE20102742}
for a Gaussian CARMA(p,q) model to the TCBm-CARMA(p,q) model.  Here we do not consider here the non-stationary
factor \(Z_t\) in equation (7) of \cite{PASCHKE20102742} but we
assume that the log price is simply CARMA(p,q) model with gaussian innovations. We highlight the fact that
extension to the ABM-CARMA(p,q) model proposed in \cite{PASCHKE20102742} is also straightforward in our context.

In \cite{PASCHKE20102742} model the futures log Price has the
following form: \[
\ln F\left(t,T\right) = \mathbf{b}^{\top}\mathcal{A}\left(t,T\right)X_t+\frac12 \mathbf{b}^{\top}\mathcal{B}\left(t,T\right)\mathbf{b}
\] where \[
\mathcal{A}\left(t,T\right)=e^{\mathbf{A}\left(T-t\right)}
\] \[
\mathcal{B}\left(t,T\right)=\int_{t}^{T}e^{\mathbf{A}\left(T-u\right)}\mathbf{e}\mathbf{e}^{\top}e^{\mathbf{A}\left(T-u\right)}\mbox{d}u.
\] If we want to evaluate a European Call Option on the Futures price,
we have to consider three points in time: time \(t\) the day where we
evaluate the contract derivative, time \(T_0>t\) the maturity of the
option contract and time \(T_F>T_0\) the maturity of the underlying
future contract. The price of the call option at time \(t\) can be
obtained using no arbitrage arguments as follows: \[
C_t=e^{-r\left(T_0-t\right)}\mathsf{E}^{\mathbb{Q}}\left[\left[F\left(T_{0},T_{F}\right)-K\right]_+\left|\mathcal{F}_t\right.\right].
\] If the state process \(\left(X_{t}\right)_{t\geq0}\) is driven by a
Brownian Motion, the price is analytic and reads as
follows: \[
C_t=e^{-r\left(T_0-t\right)}\left[F\left(t,T_F\right)\Phi\left(d_1\right)-K\Phi\left(d_2\right)\right]
\] where \[
d_{1,2}=\frac{\ln\left(\frac{F\left(t,T_F\right)}{K}\right)\pm \frac12 \sigma^2\left(t,T_0,T_F\right)}{\sigma\left(t,T_0,T_F\right)}.
\] The forward integrated variance is defined as: \[
\sigma^2\left(t,T_0,T_F\right)=\mathbf{b}^{\top}\left[\int_{t}^{T_0}e^{\mathbf{A}\left(T_F-u\right)}\mathbf{e}\mathbf{e}^{\top}e^{\mathbf{A}^\top\left(T_F-u\right)}\mbox{d}u\right]\mathbf{b}.
\] To extend in our setup this result we use the sigma field
\(\mathcal{G}_{t}^{T_F}\) therefore if the case of a TCBm-CARMA(p,q)
model we have: \[
C_t=e^{-r\left(T_0-t\right)}\mathsf{E}\left[ \mathsf{E}\left[\left(F\left(T_0,T_F\right)-K\right)_+\left|\mathcal{G}_{t}^{T_F}\right.\right]\left|\mathcal{F}_t\right.\right]
\] The internal expectation under \(\mathcal{G}_{t}^{T_F}\) is exactly
the formula in \cite{PASCHKE20102742} for a fixed value of the
integrated Variance: \[
\mathsf{E}\left[\left(F\left(T_0,T_F\right)-K\right)_+\left|\mathcal{G}_{t}^{T_F}\right.\right]=\mathsf{E}\left[\left(F\left(T_0,T_F\right)-K\right)_+\left|\mathcal{F}_{t}, \sigma^2\left(t,T_0,T_F\right)\right.\right]
\] where \[
\sigma^2\left(t,T_0,T_F\right)\left|\mathcal{G}_{t}^{T_F}\right.= \mathbf{b}^{\top}\left[\int_{t}^{T_0}e^{\mathbf{A}\left(T_F-u\right)}\mathbf{e}\mathbf{e}^{\top}e^{\mathbf{A}^\top\left(T_F-u\right)}\mbox{d}\Lambda_u\right]\mathbf{b}.
\] The conditional mean becomes: \[
\mathsf{E}\left[\left(F\left(T_0,T_F\right)-K\right)_+\left|\mathcal{F}_{t}, \sigma^2\left(t,T_0,T_F\right)=\sigma^2\right.\right]=F\left(t,T_F\right)\Phi\left(d_{1,\sigma^2}\right)-K\Phi\left(d_{2,\sigma^2}\right).
\] The Gauss-Laguerre quadrature can be used to construct the random
variable \(\sigma^2_{m,n}\left(t,T_0,T_F\right)\) following the same
approach in \eqref{eq:ApproxVarCar}. The generic \(k^{th}\) realization
of the random variable \(\sigma^2_{m,n}\left(t,T_0,T_F\right)\) has this
form: \begin{equation}
\sigma^2_{m,n,k}\left(t,T_0,T_F\right) = \underset{k=0}{\stackrel{\left[2^n\left(T_0-t\right)\right]-1}{\sum}} \mathbf{b}^\top e^{\mathbf{A}\left(T_{F}-t-k2^{-n}\right)} \mathbf{e}\mathbf{e}^{\top} e^{\mathbf{A}^\top\left(T_{F}-t-k2^{-n}\right)}\mathbf{b} u_k 
\label{eq:FIV}
\end{equation} with probability \[
\mathbb{P}\left(\sigma^2_{m,n,k}\left(t,T_0,T_F\right)\right)=   \underset{i=1}{\stackrel{m}{\prod}}\mathbb{P}^{n_i}\left(u_i\right)
\]where \(n_i\) is the times that the realization \(u_{i}\) appears in the
trajectory of the approximated subordinators and we have this
constraint: \[
\sum_{i=1}^{m}n_i=\left[2^n\left(T_0-t\right)\right]-1.
\] Now the pricing formula has the same representation in
\eqref{mainResCarmaPricing} where instead of the random variable
\(V_{t_0}^{T}\left(n,m\right)\) that can be seen as an approximation of
the spot integrated variance we have the Gauss Laguerre approximation of
the Forward Integrated Variance which realization are in
\eqref{eq:FIV}.

The same result can be applied in a straightforward manner to the case
of the European Put price when the underlying is a Future contract.
Indeed it is worth to notice the construction proposed in this paper
implies a Law convergence consequently the convergence of the formulas
in \eqref{mainResCarmaPricing} for the TCBm-CARMA(p,q) model and in
\eqref{eq:SectN1} for the Time Changed Brownian motion is ensured when
the function \(g\) is a bounded continuous function while
for a lower-semi continuous function bounded from below only a lower
bound can be established. Therefore the convergence behavior is clear
in the case of the put option prices and to avoid issues due to this
fact we perform the following steps. We first use the Gauss-Laguerre approximation scheme for the Put option price. Then we obtain the corresponding Call price using the put-call parity formula.

We report in the following Tables and figures the comparison between the
Gauss-Laguerre and MC prices for different call option prices.

\begin{figure}[!htbp]
\centering
\includegraphics[width=0.5\textwidth]{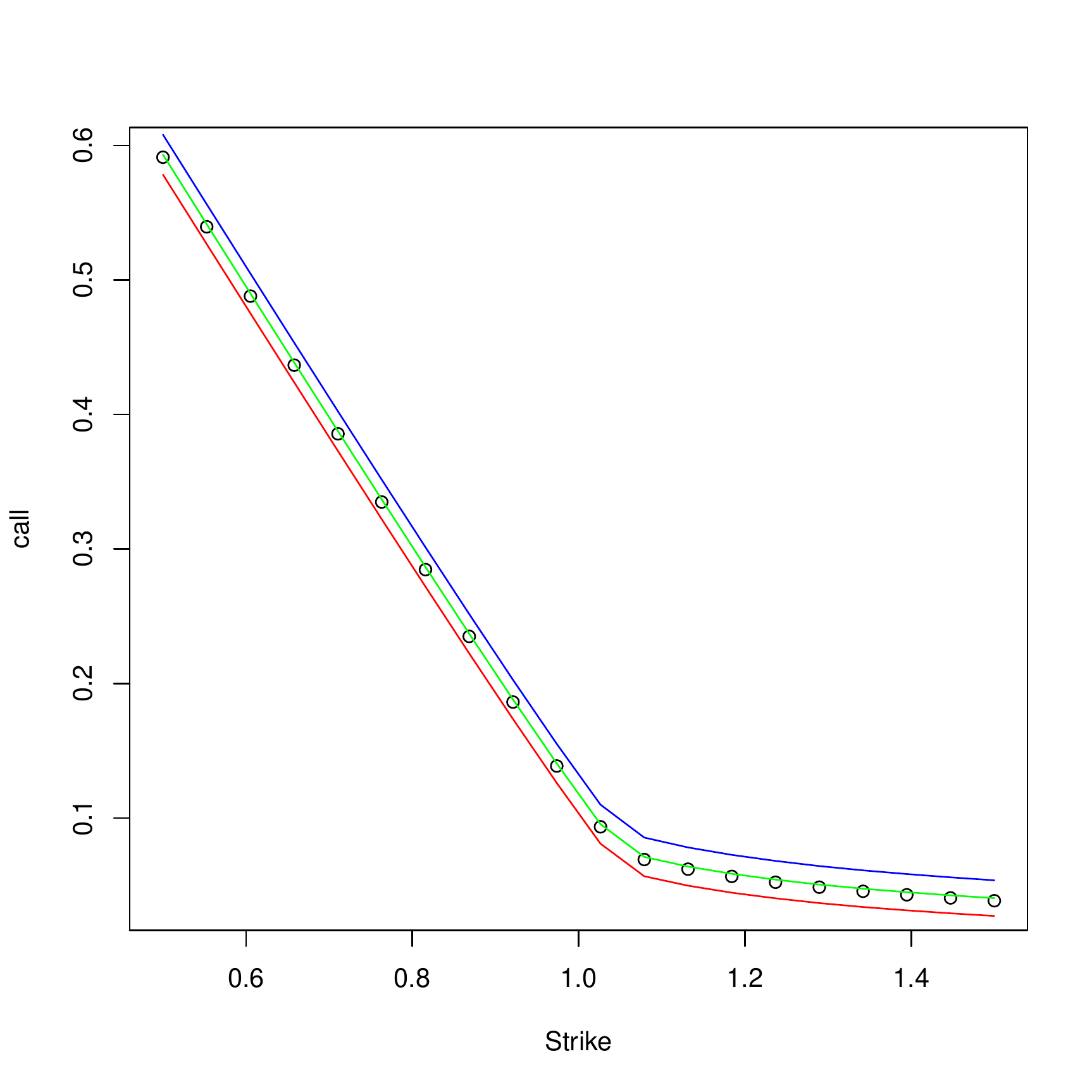}
\caption{Option Call Price with Maturity 1 Month on a Future with
maturity 2 Months}
\end{figure}

\begin{table}[!htbp] \centering  
\begin{tabular}{@{\extracolsep{5pt}} ccccc} 
\\[-1.8ex]\hline 
\hline \\[-1.8ex] 
K & Gauss L & MC & UB & LB \\ 
\hline \\[-1.8ex] 
$0.50000$ & $0.59129$ & $0.59324$ & $0.60810$ & $0.57838$ \\ 
$0.55263$ & $0.53957$ & $0.54151$ & $0.55635$ & $0.52668$ \\ 
$0.60526$ & $0.48803$ & $0.48998$ & $0.50478$ & $0.47517$ \\ 
$0.65789$ & $0.43668$ & $0.43863$ & $0.45340$ & $0.42385$ \\ 
$0.71053$ & $0.38564$ & $0.38759$ & $0.40233$ & $0.37285$ \\ 
$0.76316$ & $0.33489$ & $0.33684$ & $0.35154$ & $0.32213$ \\ 
$0.81579$ & $0.28465$ & $0.28659$ & $0.30126$ & $0.27193$ \\ 
$0.86842$ & $0.23499$ & $0.23694$ & $0.25156$ & $0.22231$ \\ 
$0.92105$ & $0.18615$ & $0.18810$ & $0.20268$ & $0.17352$ \\ 
$0.97368$ & $0.13859$ & $0.14053$ & $0.15507$ & $0.12599$ \\ 
$1.02632$ & $0.09339$ & $0.09534$ & $0.10983$ & $0.08084$ \\ 
$1.07895$ & $0.06907$ & $0.07101$ & $0.08539$ & $0.05663$ \\ 
$1.13158$ & $0.06190$ & $0.06384$ & $0.07806$ & $0.04963$ \\ 
$1.18421$ & $0.05653$ & $0.05847$ & $0.07253$ & $0.04441$ \\ 
$1.23684$ & $0.05218$ & $0.05413$ & $0.06804$ & $0.04022$ \\ 
$1.28947$ & $0.04850$ & $0.05044$ & $0.06421$ & $0.03667$ \\ 
$1.34211$ & $0.04544$ & $0.04739$ & $0.06102$ & $0.03376$ \\ 
$1.39474$ & $0.04282$ & $0.04477$ & $0.05827$ & $0.03127$ \\ 
$1.44737$ & $0.04050$ & $0.04245$ & $0.05582$ & $0.02907$ \\ 
$1.50000$ & $0.03842$ & $0.04037$ & $0.05362$ & $0.02711$ \\ 
\hline \\[-1.8ex] 
\end{tabular}
\caption{Comparison Call option prices on Futures with $T_0 = \text{1M}$ and $T_F = \text{2M}$. \label{Comp1}} 
\end{table}

\begin{figure}[!htbp]
\centering
\includegraphics[width=0.50\textwidth,height=\textheight]{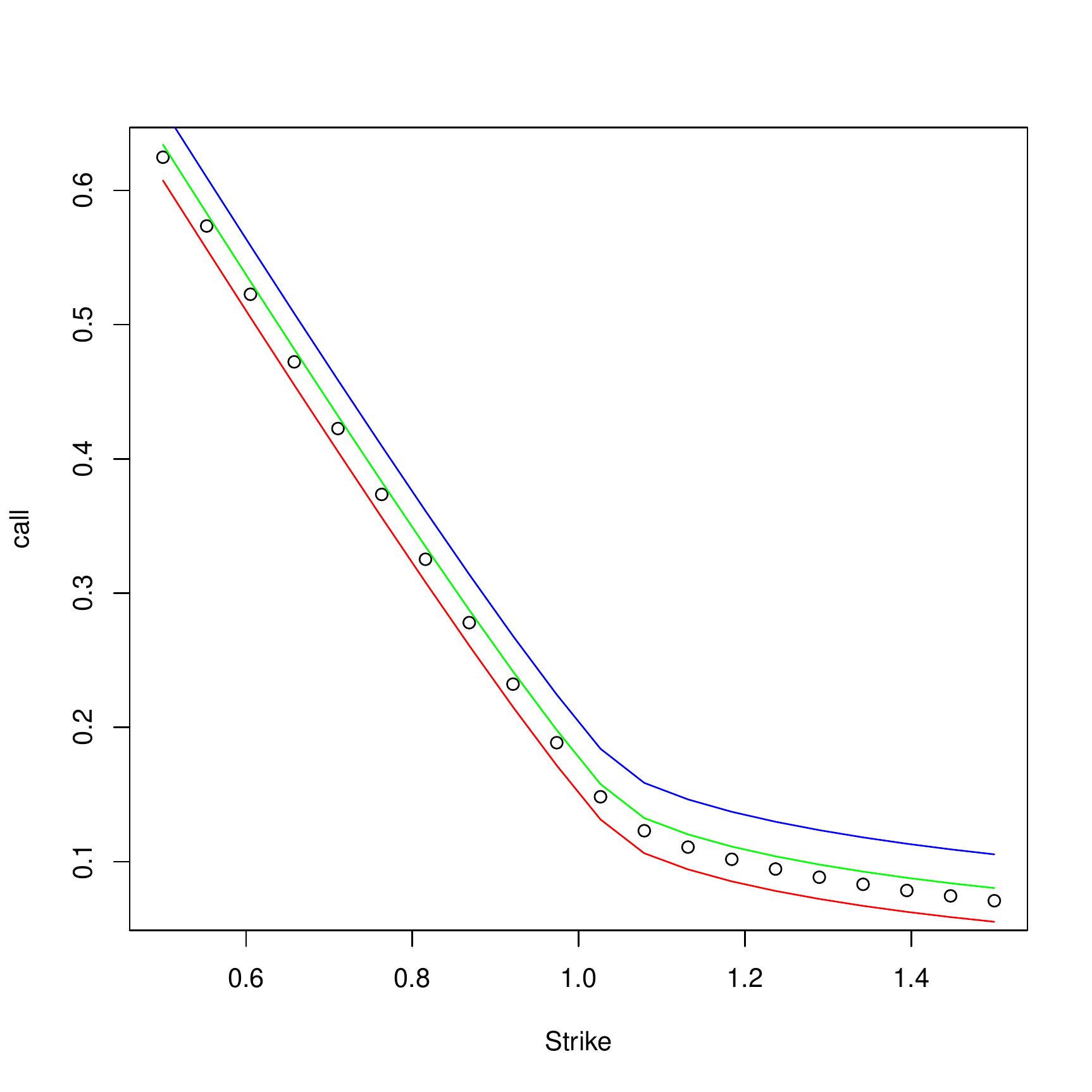}
\caption{Option Call Price with Maturity 2 Months on a Future with
maturity 3 Months}
\end{figure}

\begin{table}[!htbp] \centering 
\begin{tabular}{@{\extracolsep{5pt}} ccccc} 
\\[-1.8ex]\hline 
\hline \\[-1.8ex] 
K & Gauss L & MC & UB & LB \\ 
\hline \\[-1.8ex] 
$0.50000$ & $0.62469$ & $0.63412$ & $0.66088$ & $0.60736$ \\ 
$0.55263$ & $0.57347$ & $0.58291$ & $0.60964$ & $0.55617$ \\ 
$0.60526$ & $0.52263$ & $0.53207$ & $0.55877$ & $0.50536$ \\ 
$0.65789$ & $0.47229$ & $0.48173$ & $0.50840$ & $0.45506$ \\ 
$0.71053$ & $0.42261$ & $0.43204$ & $0.45868$ & $0.40541$ \\ 
$0.76316$ & $0.37354$ & $0.38298$ & $0.40957$ & $0.35639$ \\ 
$0.81579$ & $0.32525$ & $0.33468$ & $0.36123$ & $0.30814$ \\ 
$0.86842$ & $0.27804$ & $0.28747$ & $0.31397$ & $0.26098$ \\ 
$0.92105$ & $0.23225$ & $0.24168$ & $0.26813$ & $0.21524$ \\ 
$0.97368$ & $0.18861$ & $0.19805$ & $0.22444$ & $0.17166$ \\ 
$1.02632$ & $0.14833$ & $0.15776$ & $0.18409$ & $0.13144$ \\ 
$1.07895$ & $0.12303$ & $0.13247$ & $0.15868$ & $0.10626$ \\ 
$1.13158$ & $0.11087$ & $0.12030$ & $0.14637$ & $0.09424$ \\ 
$1.18421$ & $0.10179$ & $0.11123$ & $0.13714$ & $0.08531$ \\ 
$1.23684$ & $0.09453$ & $0.10396$ & $0.12973$ & $0.07820$ \\ 
$1.28947$ & $0.08843$ & $0.09786$ & $0.12349$ & $0.07223$ \\ 
$1.34211$ & $0.08315$ & $0.09259$ & $0.11808$ & $0.06709$ \\ 
$1.39474$ & $0.07855$ & $0.08799$ & $0.11335$ & $0.06263$ \\ 
$1.44737$ & $0.07451$ & $0.08394$ & $0.10918$ & $0.05871$ \\ 
$1.50000$ & $0.07091$ & $0.08035$ & $0.10546$ & $0.05523$ \\ 
\hline \\[-1.8ex] 
\end{tabular} 
\caption{Comparison Call option prices on Futures with $T_0 = \text{2M}$ and $T_F = \text{3M}$. \label{Comp2}} 
\end{table}

\begin{figure}[!htbp]
\centering
\includegraphics[width=0.50\textwidth,height=\textheight]{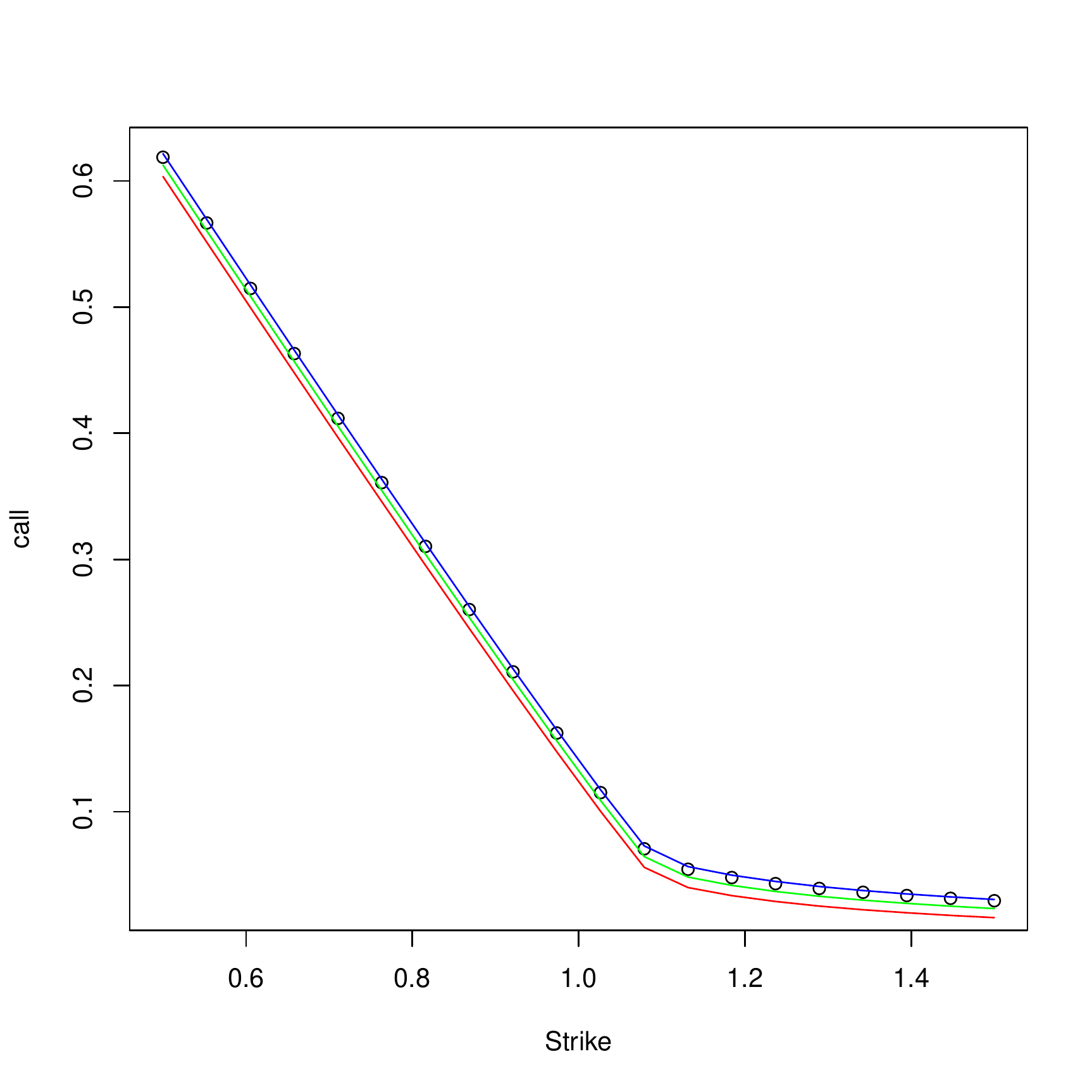}
\caption{Option Call Price with Maturity 1 month on a Future with
maturity 3 months.}
\end{figure}

\begin{table}[!htbp] \centering 
\begin{tabular}{@{\extracolsep{5pt}} ccccc} 
\\[-1.8ex]\hline 
\hline \\[-1.8ex] 
Strike & $Price^{Laguerre}$ & MC-mid & MC-lwb & MC-upb \\ 
\hline \\[-1.8ex] 
$0.50000$ & $0.61870$ & $0.61249$ & $0.62156$ & $0.60342$ \\ 
$0.55263$ & $0.56662$ & $0.56042$ & $0.56946$ & $0.55138$ \\ 
$0.60526$ & $0.51475$ & $0.50854$ & $0.51755$ & $0.49953$ \\ 
$0.65789$ & $0.46311$ & $0.45690$ & $0.46586$ & $0.44793$ \\ 
$0.71053$ & $0.41178$ & $0.40557$ & $0.41449$ & $0.39665$ \\ 
$0.76316$ & $0.36088$ & $0.35467$ & $0.36353$ & $0.34581$ \\ 
$0.81579$ & $0.31031$ & $0.30410$ & $0.31291$ & $0.29530$ \\ 
$0.86842$ & $0.26029$ & $0.25409$ & $0.26282$ & $0.24535$ \\ 
$0.92105$ & $0.21092$ & $0.20471$ & $0.21339$ & $0.19604$ \\ 
$0.97368$ & $0.16243$ & $0.15623$ & $0.16483$ & $0.14762$ \\ 
$1.02632$ & $0.11522$ & $0.10902$ & $0.11756$ & $0.10047$ \\ 
$1.07895$ & $0.07070$ & $0.06449$ & $0.07297$ & $0.05601$ \\ 
$1.13158$ & $0.05450$ & $0.04829$ & $0.05660$ & $0.03999$ \\ 
$1.18421$ & $0.04791$ & $0.04170$ & $0.04982$ & $0.03359$ \\ 
$1.23684$ & $0.04311$ & $0.03691$ & $0.04484$ & $0.02897$ \\ 
$1.28947$ & $0.03929$ & $0.03308$ & $0.04084$ & $0.02531$ \\ 
$1.34211$ & $0.03623$ & $0.03002$ & $0.03763$ & $0.02241$ \\ 
$1.39474$ & $0.03369$ & $0.02748$ & $0.03494$ & $0.02001$ \\ 
$1.44737$ & $0.03147$ & $0.02526$ & $0.03259$ & $0.01794$ \\ 
$1.50000$ & $0.02956$ & $0.02335$ & $0.03054$ & $0.01615$ \\ 
\hline \\[-1.8ex] 
\end{tabular}
  \caption{Comparison of Call option prices on Futures with $T_0 = \text{1 month}$ and $T_F = \text{3 months}$. We compute the price using the approximation procedure ($Price^{Laguerre}$) and compare it with Monte Carlo prices (MC-mid, MC-lwb is the 5\% quantile of MC simulations while MC-upb is the 5\% quantile of MC simulations).  }
  \label{Comp3} 
\end{table}
\clearpage
\section{Conclusion}\label{conclusion}
In this paper we propose an approximation procedure for the evaluation of the transition density of a TCBm-CARMA(p,q) process that resultsto be a finite mixture of normals. Exploiting this structure we obtain a simple estimation procedure and  pricing formulas for financial contracts whose value depend only on the value of the underlying at maturity modelled as an exponential TCBm-CARMA(p,q). A possible extension of our proposed approximation methodology to the pricing of path dependent contracts may be based on the result in \cite{Hieber2012} for the evaluation of the first passage time for a Time Changed Brownian Motion. Indeed the  process $V_{t_0}^{t}$ has the same structure of a subordinator while the TCBm-CARMA can be seen as a TCBm where the random time is the process $V_{t_0}^{t}$. This could  also give us the possibility to extend our approach to the evaluation of the density function for the time-until death variable that is necessary for the evaluation of contracts with minimimum guaranteed death benefit.


\section{Appendix}\label{appendix}

\subsection{EM algorithm}
\label{EMderivation}
We derive the Expectation Maximization algorithm for the approximated
density in \eqref{eq:approximated}. As a first step we determine the
complete-data log-likelihood function defined as: \begin{eqnarray}
\mathcal{L}^{\star}\left(\mu_0,\mu, \sigma, \varphi_+, \lambda,\theta\right)&=&\sum_{t=1}^T\ln\left[\phi\left(y_t;\mu_0+\mu U_t; \sigma^2 U_t\right)\mathbb{P}\left(U_t,\varphi_+,\lambda,\theta\right)\right]\nonumber\\
&=&\sum_{t=1}^T\sum_{i=1}^{m}D_{t,i}\ln\left[\phi\left(y_t;\mu_0+\mu u_i; \sigma^2 u_i\right)\mathbb{P}\left(u_i,\varphi_+,\lambda,\theta\right)\right]
\end{eqnarray} where \(D_{t,i}\) assumes value 1 when \(U_t=u_i\) and
\(0\) otherwise. Following the seminal work of \cite{Dempster77maximumlikelihood}, we perform the Expectation-step (E-step henceforth)
evaluating the conditional distribution of the variables
\(\left\{U_t\right\}_{t=1,\ldots,T}\) given the observed data. Applying
the Bayes' theorem we have: \[
\mathbb{P}\left(U_t=u_i\left|y_t,\Theta_{h-1}\right.\right)=\frac{\phi\left(y_t;\mu_{0,h-1}+\mu_{h-1} u_i; \sigma^2_{h-1} u_i\right)\mathbb{P}\left(u_i,\varphi_{+,h-1},\lambda_{h-1},\theta_{h-1}\right)}{\underset{i=1}{\stackrel{m}{\sum}}\phi\left(y_t;\mu_{0,h-1}+\mu_{h-1} u_i; \sigma^2_{h-1} u_i\right)\mathbb{P}\left(u_i,\varphi_{+,h-1},\lambda_{h-1},\theta_{h-1}\right)}
\] where
\(\Theta_{h-1} =\left(\mu_{0,h-1},\mu_{h-1}, \sigma^2_{h-1},\varphi_{+,h-1}, \lambda_{h-1},\theta_{h-1}\right)\).
The E-step consists of computing the conditional expectation of
\(\mathcal{L}^{\star}\left(\mu_0,\mu,\varphi_+, \lambda,\theta\right)\)
in the following way: \[
\mathbb{E}\left[\mathcal{L}^{\star}\left(\mu_{0,h},\mu_{h}, \sigma_{h},\varphi_{+,h}, \lambda_{h},\theta_{h}\right)\right]=\sum_{i=1}^{m}\sum_{t=1}^T\ln\left[\phi\left(y_t;\mu_{0,h}+\mu_{h} u_i; \sigma^2_{h} u_i\right)\mathbb{P}\left(u_i,\varphi_{+,h},\lambda_{h},\theta_{h}\right)\right]\mathbb{P}\left(U_t=u_i\left|y_t,\Theta_{h-1}\right.\right).
\] Recalling that \(u_i=\frac{k_i}{\varphi_+}\) we get: \footnotesize
\begin{eqnarray}
\mathbb{E}\left[\mathcal{L}^{\star}\left(\mu_{0,h},\mu_{h},\sigma_{h},\varphi_{+,h}, \lambda,\theta\right)\right]&=&\sum_{i=1}^{m}\sum_{t=1}^T\ln\left[\phi\left(y_t;\mu_{0,h}+\mu_{h} \frac{k_i}{\varphi_{+,h}}; \sigma^2_h\frac{k_i}{\varphi_{+,h}}\right)\mathbb{P}\left(\frac{k_i}{\varphi_{+,h}},\varphi_{+,h},\lambda_{h},\theta_{h}\right)\right]\mathbb{P}\left(U_t=\frac{k_i}{\varphi_{+,h-1}}\left|y_t,\Theta_{h-1}\right.\right)\nonumber\\
&=&\sum_{i=1}^{m}\sum_{t=1}^T\ln\left[\phi\left(y_t;\mu_{0,h}+\mu_{h} \frac{k_i}{\varphi_{+,h}}; \sigma^2_h\frac{k_i}{\varphi_{+,h}}\right)\right]\mathbb{P}\left(U_t=\frac{k_i}{\varphi_{+,h-1}}\left|y_t,\Theta_{h-1}\right.\right)\nonumber\\
&+&\sum_{i=1}^{m}\sum_{t=1}^T\ln\left[\mathbb{P}\left(\frac{k_i}{\varphi_{+,h}},\varphi_{+,h},\lambda_{h},\theta_{h}\right)\right]\mathbb{P}\left(U_t=\frac{k_i}{\varphi_{+,h-1}}\left|y_t,\Theta_{h-1}\right.\right)
\label{quant2}
\end{eqnarray} \normalsize The Maximization-step (M-step henceforth) is
based on the maximization of the quantity in \eqref{quant2}, i.e.:
\begin{equation}
\left(\mu_{0,h},\mu_{h},\sigma_{h},\varphi_{+,h}, \lambda_{h},\theta_{h}\right)= \underset{
\begin{array}{c} 
\scriptsize{\mu_{0,h},\mu_{h},\sigma_{h}} \\ 
\scriptsize{\varphi_{+,h}, \lambda_{h},\theta_{h}}
\end{array}
}{\text{argmax }}   \mathbb{E}\left[\mathcal{L}^{\star}\left(\mu_{0,h},\mu_{h},\sigma_{h},\varphi_{+,h}, \lambda_{h},\theta_{h}\right)\right]
\label{Prob1}
\end{equation} Using the following parametrization: \[
\left\{ 
\begin{array}{l}
\mu=\tilde{\mu}\varphi_+\\
\sigma=\tilde{\sigma}\sqrt{\varphi_+}
\end{array}
\right. .
\] The problem in \eqref{Prob1} becomes: \tiny \[
 \underset{
\begin{array}{c} 
\scriptsize{\mu_{0,h},\mu_{h},\sigma_{h}} \\ 
\scriptsize{\varphi_{+,h}, \lambda_{h},\theta_{h}}
\end{array}
}{\text{argmax }} \sum_{i=1}^{m}\sum_{t=1}^T\ln\left[\phi\left(y_t;\mu_{0,h}+\tilde{\mu}_{h} k_i; \tilde{\sigma}^2_h k_i\right)\right]\mathbb{P}\left(U_t=\frac{k_i}{\varphi_{+,h-1}}\left|y_t,\Theta_{h-1}\right.\right) + \sum_{i=1}^{m}\sum_{t=1}^T\ln\left[\mathbb{P}\left(\frac{k_i}{\varphi_{+,h}},\varphi_{+,h},\lambda_{h},\theta_{h}\right)\right]\mathbb{P}\left(U_t=\frac{k_i}{\varphi_{+,h-1}}\left|y_t,\Theta_{h-1}\right.\right)  
\] \normalsize that can be split as follows: \begin{equation}
\underset{
\scriptsize{\mu_{0,h},\mu_{h},\sigma_{h}} 
}{\text{argmax }} \mathbb{H}_1\left(\mu_{0,h},\mu_{h},\sigma_{h}\right):=\sum_{i=1}^{m}\sum_{t=1}^T\ln\left[\phi\left(y_t;\mu_{0,h}+\tilde{\mu}_{h} k_i; \tilde{\sigma}^2_h k_i\right)\right]\mathbb{P}\left(U_t=\frac{k_i}{\varphi_{+,h-1}}\left|y_t,\Theta_{h-1}\right.\right) 
\label{Prob:1a}
\end{equation} \begin{equation}
\underset{
\scriptsize{\varphi_{+,h}, \lambda_{h},\theta_{h}}
}{\text{argmax }}  \mathbb{H}_2\left(\varphi_{+,h}, \lambda_{h},\theta_{h}\right):=\sum_{i=1}^{m}\sum_{t=1}^T\ln\left[\mathbb{P}\left(\frac{k_i}{\varphi_{+,h}},\varphi_{+,h},\lambda_{h},\theta_{h}\right)\right]\mathbb{P}\left(U_t=\frac{k_i}{\varphi_{+,h-1}}\left|y_t,\Theta_{h-1}\right.\right)
\label{Prob:2a}
\end{equation}

\subsection{Gauss Laguerre Quadrature}\label{gauss-laguerre-quadrature}

In this section we review some results about the Gauss-Laguerre 
quadrature  necessary to understand the behavior of
our approximation scheme. We refer to \cite{Rabinowitz1967,Uspensky1928,abramowitz70a} for a complete discussion about this  quadrature.

\noindent Let \(f\left(x\right)\) be a continuous function on the support
\(\left[0,+\infty\right)\) and let the integral
\(\int_{0}^{+\infty}f\left(x\right)e^{-x}\mbox{d}x<+\infty\) be finite
with \(f\) be \(2m\) differentiable. Then we have: \[
\int_{0}^{+\infty}e^{-x}f\left(x\right)\mbox{d}x=\sum_{i=1}^m\omega\left(u_i\right)f\left(u_i\right)+ \mathcal{R}_{m}
\] where \[
\mathcal{R}_{m}=\frac{\left(m!\right)^2}{\left(2m\right)!}f^{\left(2m\right)}\left(\epsilon\right), \ \ \epsilon \in (0,+\infty).
\]

\subsection{Generalized Gauss Laguerre Quadrature}\label{generalized-gauss-laguerre-quadrature}

The Generalized
Gauss-Laguerre quadrature can be applied in the presence of non negligible
singularity at \(x=0\). Following \cite{Rabinowitz1967}, let \(f\left(x\right)\) be a non-negative
continuous function such that \(\omega\left(x\right)f\left(x\right)\) is
a monotonically non negative not increasing in \(\left(0,+\infty\right)\)
where \(\omega\left(x\right)= x^{\alpha}e^{-x}, \ \alpha>-1\), \[
f\left(x\right)\leq \frac{e^{x}}{x^{\alpha+1+\rho}}
\] for some \(\rho>0\) then, if the function \(f\left(x\right)\) is
\(2n\) differentiable, the Generalized Gauss-Laguerre quadrature has the
following form: \[
\int_{0}^{+\infty} \omega\left(x\right)f\left(x\right)\mbox{d}x=\sum_{i=1}^{m}\omega\left(u_i\right)f\left(u_i\right)+\mathcal{R}_{m},
\] with
\(\omega\left(u_i\right)=\frac{\Gamma\left(m+\alpha\right)u_i}{m!\left(m+\alpha\right)\left[L^{\alpha}_{m-1}\left(u_i\right)\right]^2}\)
and \(L_m^{\alpha}\left(x\right)\) is the generalized Laguerre polynomial.

The residual term \(\mathcal{R}_{m}\) can be written as: \[
\mathcal{R}_m=\frac{m!\Gamma\left(m+\alpha+1\right)}{\left(2m\right)!}f^{\left(2m\right)}\left(\epsilon\right),\ \epsilon\in\left(0,+\infty\right).
\]

A standard example where it is necessary to use the Generalized
Gauss-Laguerre quadrature is the numerical evaluation of the moment
generating function of a Gamma random variable with shape parameter
\(\alpha \in \left(0,1\right)\). The usage of the Generalized Gauss
Laguerre is due to the fact that, in this case, we have a singularity at
\(x=0\); the requirements described in this section can be easily
checked and the error term can be evaluated due to smooth condition of
the exponential function. For the case of \(\alpha \geq 1\) the standard
Gauss Laguerre quadrature described in the previous section can be
easily applied.

\subsection{Error computation in the option pricing formula in the case of NVMM}
\label{erroroptNVMM}

It is worth to notice that the
formula in \eqref{eq:SectN1} can be written as: \[
\sum_{i=1}^{m}\mathsf{E}\left[g\left(\mu+\theta \Lambda_m+\sqrt{\Lambda_m}Z\right)\left|\mathcal{F}_0,\Lambda_m=u_i\right.\right]\mathbb{P}\left(u_i\right)=\frac{A_m}{B_m}
\] where: \[
A_m=\sum_{i=1}^{m}\mathsf{E}\left[g\left(\mu+\theta \Lambda_m+\sqrt{\Lambda_m}Z\right)\left|\mathcal{F}_0,\Lambda_m=u_i\right.\right]\frac{\omega\left(k_i\right)}{k_i}\left(\frac{k_i}{\varphi_+}\right)^{\lambda}L_{\theta}\left(\frac{k_i}{\varphi_+}\right), \ k_i=u_i\varphi_+
\] and \[
B_m=\sum_{j=1}^m\frac{\omega\left(k_j\right)}{k_{j}}\left(\frac{k_j}{\varphi_+}\right)^{\lambda}L_{\theta}\left(\frac{k_j}{\varphi_+}\right).
\] We analyze the term \(A_m\) as \(m\rightarrow+\infty\), by Gauss -
Laguerre Quadrature we have: \begin{equation}
\lim_{m\rightarrow +\infty}A_m=\int_{0}^{+\infty}\mathsf{E}\left[g\left(\mu \Lambda +\theta \Lambda_T+\sqrt{\Lambda_T}Z \right)\left|\mathcal{F}_0, \Lambda_T=k\right.\right]\left(\frac{k}{\varphi_+}\right)^{\lambda}\frac{L_{\theta}\left(k/\varphi_+\right)}{k}\mbox{d}k,
\label{limit:A_n}
\end{equation} where the integral in the right hand is exactly the expectation
of the function \(g\left(Y_{T}\right)\) where \(Y_T\) is a normal
variance mean mixture (it is enough to solve the integral using the
substitution \(\frac{k}{\varphi_+}=u\)). Denoting with \(A\) the
integral in \eqref{limit:A_n}, we have the following result due to the
standard Gauss-Laguerre quadrature: \[
A= A_m+\mathcal{R}_{m}\left(A_m\right)
\] where the remaining term has the following form: \[
\mathcal{R}_{m}\left(A_m\right)= \frac{\left(m!\right)^2}{\left(2m\right)!} \partial^{2m}\left[\mathsf{E}\left[g\left(\mu T +\theta \Lambda_T+\sqrt{\Lambda_T}Z \right)\left|\mathcal{F}_0, \Lambda_T=\epsilon\right.\right]\left(\frac{\epsilon}{\varphi_+}\right)^{\lambda}\frac{L_{\theta}\left(\epsilon/\varphi_+\right)}{\epsilon}\right], \ \epsilon \in\left(0,+\infty\right).
\] 
A discussion about the behaviour of the remaining term $\mathcal{R}_{m}\left(A_m\right)$ can be found in \cite{lubinsky1983geometric}. The author proved, under mild conditions, the geometric convergence for a Gauss-Laguerre quadrature for a function that can be written as a power series [see \cite{Mastroianni1994,XIANG2012434} for a complete discussion and generalizations]. 

We analyze the behaviour of term \(B_m\) that: \begin{equation}
\lim_{m\rightarrow+\infty}B_m=\int_0^{+\infty}\left(\frac{k}{\varphi_+}\right)^{\lambda}\frac{L_{\theta}\left(k/\varphi_+\right)}{k}\mbox{d}k.
\label{limit:B_n}
\end{equation} Using the substitution \(u=\frac{k}{\varphi_+}\), the
integral is equal to one because the integrand function is the density
in \eqref{2}. Denoting with \(B\) the integral in \eqref{limit:B_n} we
have \[
B = B_m + \mathcal{R}_{m}\left(B_m\right).
\] The remaining term \(\mathcal{R}_{m}\left(B_m\right)\) has the
following form: \[
\mathcal{R}_{m}\left(B_m\right)=\frac{\left(m!\right)^2}{\left(2m\right)!}\partial^{2m}\left[\left(\frac{\epsilon}{\varphi_+}\right)^{\lambda}\frac{L_{\theta}\left(\epsilon/\varphi_+\right)}{\epsilon}\right], \ \epsilon \in \left(0,+\infty\right)
\] 

We are now able to establish the error term behaviour of our
approximation approach for the normal variance mean mixture. The result
presented here holds when we have a no negligible singularity at \(x=0\)
but the result for this type approximation can easily to generalize to
case of the singularity at \(x=0\) using the Generalized Gauss-Laguerre
quadrature.

We define the error term \(\mathcal{R}_m\) as: \begin{eqnarray}
\mathcal{R}_m &:=& \mathsf{E}\left[g\left(Y_T\right)\left|\mathcal{F}_0\right.\right]-\mathsf{E}\left[g\left(Y^{m}_T\right)\left|\mathcal{F}_0\right.\right] \nonumber \\
&=& \frac{\mathcal{A}_m+\mathcal{R}_{m}\left(\mathcal{A}_m\right)}{\mathcal{B}_m+\mathcal{R}_{m}\left(\mathcal{B}_m\right)}-\frac{\mathcal{A}_m}{\mathcal{B}_m} \nonumber \\
&=&\frac{\mathcal{A}_m+\mathcal{R}_{m}\left(\mathcal{A}_m\right)}{\mathcal{B}_m+\mathcal{R}_{m}\left(\mathcal{B}_m\right)}-\frac{\mathcal{A}_m}{\mathcal{B}_m+\mathcal{R}_{m}\left(\mathcal{B}_m\right)}+\frac{\mathcal{A}_m}{\mathcal{B}_m+\mathcal{R}_{m}\left(\mathcal{B}_m\right)}-\frac{\mathcal{A}_m}{\mathcal{B}_m} \nonumber \\
\end{eqnarray} Noting that
\(\mathcal{R}_{m}\left(\mathcal{B}_m\right)+\mathcal{B}_m=1\), we have
\[
\mathcal{R}_m = \mathcal{R}_{m}\left(\mathcal{A}_m\right)-\frac{\mathcal{A}_m}{\mathcal{B}_m}\mathcal{R}_{m}\left(\mathcal{B}_m\right)
\] Therefore \[
\left|\mathcal{R}_m\right|\leq \left|\mathcal{R}_{m}\left(\mathcal{A}_m\right)\right|+\left|\frac{\mathcal{A}_m}{\mathcal{B}_m}\right|\left|\mathcal{R}_{m}\left(\mathcal{B}_m\right)\right|.
\]

\end{document}